\documentclass[preprint]{aastex}

\received{}
\accepted{}
\journalid{}{}
\articleid{}{}

\shortauthors{Carpenter et al.}
\shorttitle{Variable Stars in Chamaeleon~I}

\newcommand{\ts}{\thinspace}
\newcommand{\simless}{\mathbin{\lower 3pt\hbox
     {$\rlap{\raise 5pt\hbox{$\char'074$}}\mathchar"7218$}}}
\newcommand{\simgreat}{\mathbin{\lower 3pt\hbox
     {$\rlap{\raise 5pt\hbox{$\char'076$}}\mathchar"7218$}}}

\newcommand{\about}    {$\sim$\ts}
\newcommand{\aboutless}{$\simless$\ts}
\newcommand{\aboutmore}{$\simgreat$\ts}

\newcommand{\KB}{$K_s$}
\newcommand{\HK}{$H-K_s$}

\newcommand{\msun}{\ts M$_\odot$}
\newcommand{\kms}{\ts km\ts s$^{-1}$}

\newcommand{\etal}{et~al.}

\def\insertplot#1#2#3#4#5#6#7{
\vskip 10pt\nobreak\hbox to \hsize{\hss\dimen0=#3in\hbox to #6\dimen0{%
\dimen0=#2in\vbox to #6\dimen0{\vss
\special{ps: plotfile #1}
\special{ps::[end]
  PGPLOT restore
}
}\hss}\hss}\vskip 10pt}

\begin{document}

\title{Near-Infrared Photometric Variability of\\
       Stars Toward the Chamaeleon~I Molecular Cloud}

\author{John M. Carpenter, Lynne A. Hillenbrand}
\affil{California Institute of Technology, 
       Department of Astronomy, MS 105-24, \\ Pasadena, CA 91125}
\email{jmc@astro.caltech.edu, lah@astro.caltech.edu}

\author{M. F. Skrutskie\altaffilmark{1}}
\affil{University of Massachusetts, Department of Astronomy, \\
       Amherst, MA 01003}
\email{mfs4n@virginia.edu}
\altaffiltext{1}{Current Address: University of Virginia, Department of 
                 Astronomy, P.O. Box 3818, Charlottesville, VA 22903}

\and

\author{Michael R. Meyer}
\affil{University of Arizona, Steward Observatory,
       \\ Tucson, AZ 85721}
\email{mmeyer@as.arizona.edu}

\begin{abstract}

We present the results of a $J$, $H$, and $K_s$ photometric monitoring 
campaign of a $0.72^\circ\times6^\circ$ area centered on the Chamaeleon~I star 
forming region. Data were obtained on 15 separate nights over a 4 month time 
interval using the 2MASS South telescope. Out of a total of 34,539 sources 
brighter than the photometric completeness limits ($J$=16.0, $H$=15.2, 
$K_s$=14.8), 95 exhibit near-infrared variability in one or more bands. The 
variables can be grouped into a population of bright, red objects that 
are associated with the Chamaeleon~I association, and a population of faint, 
blue variables that are dispersed over the full 6\arcdeg\ of the survey and 
are likely field stars or older pre-main-sequence stars unrelated to the 
present-day Chamaeleon~I molecular cloud. Ten new candidate members of 
Chamaeleon~I, including 8 brown dwarf candidates, have been identified based on 
variability and/or near-infrared excess emission in the $J-H$ vs. $H-K_s$ 
color-color-diagram. We also provide a compendium of astrometry and $J$, $H$, 
and $K_s$ photometry for previously identified members and candidate members of 
Chamaeleon~I.

\end{abstract}

\keywords{infrared: stars --- 
          stars:pre-main-sequence --- 
          stars:variables ---\hfill\\
          open clusters and associations}

\section{Introduction}

Near-infrared variability provides a means to identify young stellar 
populations independent of most current observational selection techniques 
(e.g. H$\alpha$, Li, x-ray, and near-infrared excess surveys), and has been 
shown to be sensitive to stars both with and without circumstellar accretion 
disks \citep{Skrutskie96}. Variability studies, therefore, help provide a more 
complete census of the stellar population in star forming regions. In the 
first (Carpenter, Hillenbrand, \& Skrutskie~2001, hereafter CHS01) of a series 
of papers on the near-infrared photometric variability properties of 
pre-main-sequence objects, we analyzed $J$, $H$, and $K_s$ time-series data of 
nearly 18,000 stars distributed over a $0.84^\circ\times6^\circ$ region toward 
the Orion~A molecular cloud using observations conducted with southern 2MASS 
telescope. The vast majority of the $\sim$1200 stars with time-variable 
JHK$_{\rm s}$ photometry identified in that study are young pre-main sequence 
stars associated with the Orion molecular cloud. The large sample of variables 
was used to investigate the characteristics and origins of near-infrared 
variability in young stars. 

A diversity of photometric behavior was observed during the \about 30-day time
period encompassed by the Orion observations, including cyclic fluctuations, 
eclipses, aperiodic short-term fluctuations, slow drifts in brightness over 
the full length of the observing period, colorless variability, stars that 
become redder as they fade, and stars that become bluer as they fade. 
Rotational modulation by cool spots alone can explain the observed variability 
characteristics in \about 56-77\% of the stars, while the properties of the 
photometric fluctuations are more consistent with hot spots or extinction 
changes in at least 23\% of the stars, and with variations in the mass 
accretion rate or inner radius changes in the disk in \about 1\% of the stars. 

One limitation of the Orion survey is that at the distance of the Orion 
molecular cloud, the flux-limited observations were most sensitive to 
variability in stars more massive than 1\msun. Observations of more proximate 
regions can use variability as a probe of lower mass stars and substellar 
objects and establish if the variability characteristics vary as a function
of mass. Accordingly, we have conducted a near-infrared monitoring 
campaign of the Chamaeleon~I star forming region. At a distance of 160~pc 
\citep{Whittet97}, Chamaeleon~I is three times closer than the Orion molecular 
cloud, and contains a moderately large sample of young pre-main sequence stars 
(\about 120) that have already been identified (see Lawson, Feigelson, \& 
Huenemoerder 1996 and references therein). Using the 2MASS south telescope, we 
observed a $0.72^\circ\times6^\circ$ region centered on the Chamaeleon~I 
molecular cloud on 15 nights over a 4 month time period. These observations 
monitored over 34,000 stars with sensitivity to detect variability in young 
brown dwarfs with masses as low as \about 0.05\msun.

The observations and data reduction procedures for this data set are described 
in Section~\ref{data}. In Section~\ref{variable_stars}, we identify the 
variable stars from the time series photometry, present light curves for all 
of the variable stars, and examine the spatial distribution and near-infrared 
photometric characteristics of the variable population. In 
Section~\ref{candidates}, we identify candidate members of Chamaeleon~I based 
on near-infrared variability and also near-infrared excesses inferred from the 
time-averaged photometry. Our conclusions are summarized in 
Section~\ref{summary}. A list of near-infrared photometry of known and 
candidate members of Chamaeleon~I is provided in the Appendix.

\section{Data}
\label{data}

\subsection{Observations and Data Processing}
\label{observations}

The observing and data reduction procedures for the Chamaeleon~I data closely
follow that used in the Orion variability study. This section summarizes 
aspects of the data relevant specifically to the Chamaeleon~I data. A more
complete description of the data processing procedures can be found in
CHS01 and \citet{Cutri00}.

The $J$, $H$, and \KB\ observations of Chamaeleon~I were obtained with the 
2MASS 1.3 meter telescope at Cerro Tololo, Chile near the completion of the 
southern survey operations when auxiliary projects were scheduled in otherwise 
idle telescope time. All data were collected in standard 2MASS observing mode 
by scanning the telescope in declination to cover tiles of size 
($\Delta\alpha\times\Delta\delta$) \about ($8.5'\times6^\circ$) in the three 
bands simultaneously. The nominal region surveyed for this project consists of 
eight contiguous tiles in right ascension as listed in Table~\ref{tbl:coords}, 
with each tile centered on declination $-77$\arcdeg. The total sky coverage is 
\about $0.72^\circ \times 6^\circ$. Observations were obtained on 15 separate 
nights, 13 of these in April and May, 2000, and 2 in January 2000 when 
Chamaeleon~I was observed as part of the normal 2MASS survey operations
covering the entire sky. Complete coverage of all 8 tiles was obtained on
9 separate nights, while scheduling constraints limited observations to a 
subset of the eight tiles on the other 6 nights.

Adjacent tiles overlap by at least 40\arcsec\ in the 2MASS observing procedure 
to ensure complete coverage of the sky as well as a check on the photometric
repeatability of the observations. In practice, since the telescope is 
stepped slightly in right ascension while scanning a tile in declination, the 
amount of right ascension overlap changes with declination. For low 
declination regions such as Chamaeleon~I, there is near complete overlap at 
the southernmost declinations within a tile in order to maintain the minimum 
overlap in the north.  Thus for a large fraction of the sources there is more 
than one measurement per night.  Of the sources meeting the photometric 
completeness limits described in the following section, 53\% have 12-14 
photometric measurements, 38\% of have 23-25 measurements, and the rest mainly 
have intermediate number of measurements due to photometric incompleteness and 
nights with partial coverages.

The data were reduced using a development version of the 2MASS data processing 
pipeline at the Infrared Processing and Analysis Center (IPAC) that also
generated the data products for the 2MASS first and second incremental release 
catalogs. The data discussed in this paper though were not part of these 
incremental releases and will ultimately be replaced by the results of the 
final 2MASS processing. The 2MASS Explanatory Supplement \citep{Cutri00} 
contains complete details of the IPAC data reduction procedures. As with the 
Orion data, the photometric zero point was adjusted by us for each tile using 
bright, isolated stars as secondary standards (see CHS01 for full description 
of the methods). On a few scans, the 
photometric offsets derived from the secondary standards deviated 
systematically from the mean offset by up to 0.1-0.2 magnitudes over a 
$\Delta\delta$ \about 0.5\arcdeg\ region of the 6$\arcdeg$ long tile. These 
photometric deviations are presumably caused by clouds passing overhead.
For the affected region within a scan, the photometric offsets as a function 
of declination were computed by averaging the offsets for every 
10-15 secondary standards. Photometric offsets were derived in this manner for 
portions of 13 scans out of a total of 108 scans observed for this project, 
and affected only \about 1\% of the total scan area.

\subsection{Point Source List}
\label{sample}

As a first step in generating the point source list, we established the
photometric completeness limit of the observations. Using data from the 9 
nights on which all eight tiles were observed (see Table~\ref{tbl:coords}), we 
empirically determined the magnitude limit at which a star is expected to be 
detected on at least 8 of the 9 nights in the absence of source confusion.
The lack of a detection on the one night can be attributed typically to 
random noise that puts the star below the sensitivity limit of the 
observations. The completeness limit as defined here then is the magnitude at 
which there is a 89\% chance that the star was detected on an appropriate 
apparition. This magnitude limit occurs at approximately $J$=16.0, $H$=15.2, 
and $K_s$=14.8 for these data, corresponding to a signal-to-noise ratio of 
\about 7 as discussed below. 

Our initial point source list, generated from the 9 nights in which all tiles
were observed, contained 34,539 sources each having an average magnitude 
brighter than or equal to the photometric completeness limit in at least one 
band. Of these, \about 96\% have no artifact or confusion flags from the 
processing pipeline in any of the observations. After removing those sources 
flagged as persistence or filter glints, potential lingering artifacts were 
identified as objects that had unusually blue colors for stars, were detected 
less than 8 times, or had flags indicating contaminated or confused photometry 
from a nearby star.  Many of these 1266 sources were visually inspected in the 
images, and sources deemed as artifacts were then removed from the point 
source list.  Criteria were established from this exercise to identify and 
remove likely artifacts for those sources not examined individually. Of the 
34,539 stars meeting the magnitude completeness criteria, only 185 were deemed 
artifacts. The final source list for our variability analysis therefore 
contains 34,354 stars brighter than the defined completeness limits ($J$=16.0, 
$H$=15.2, \KB=14.8) in at least one band. Compared to the Orion observations
(CHS01), Chamaeleon~I contains nearly twice the number of sources 
(34,354 vs. 18,552) since it is both closer to the galactic center 
($l=297$\arcdeg\ vs. 209\arcdeg) and the galactic plane ($b=-16$\arcdeg\ vs. 
$-19$\arcdeg). However, the number of sources removed as artifacts is 4 times 
lower (185 vs. 744) since Chamaeleon~I does not have the combination of high 
stellar density and bright nebulosity that makes photometry and source 
identification difficult in Orion.

\begin{figure}
\insertplot{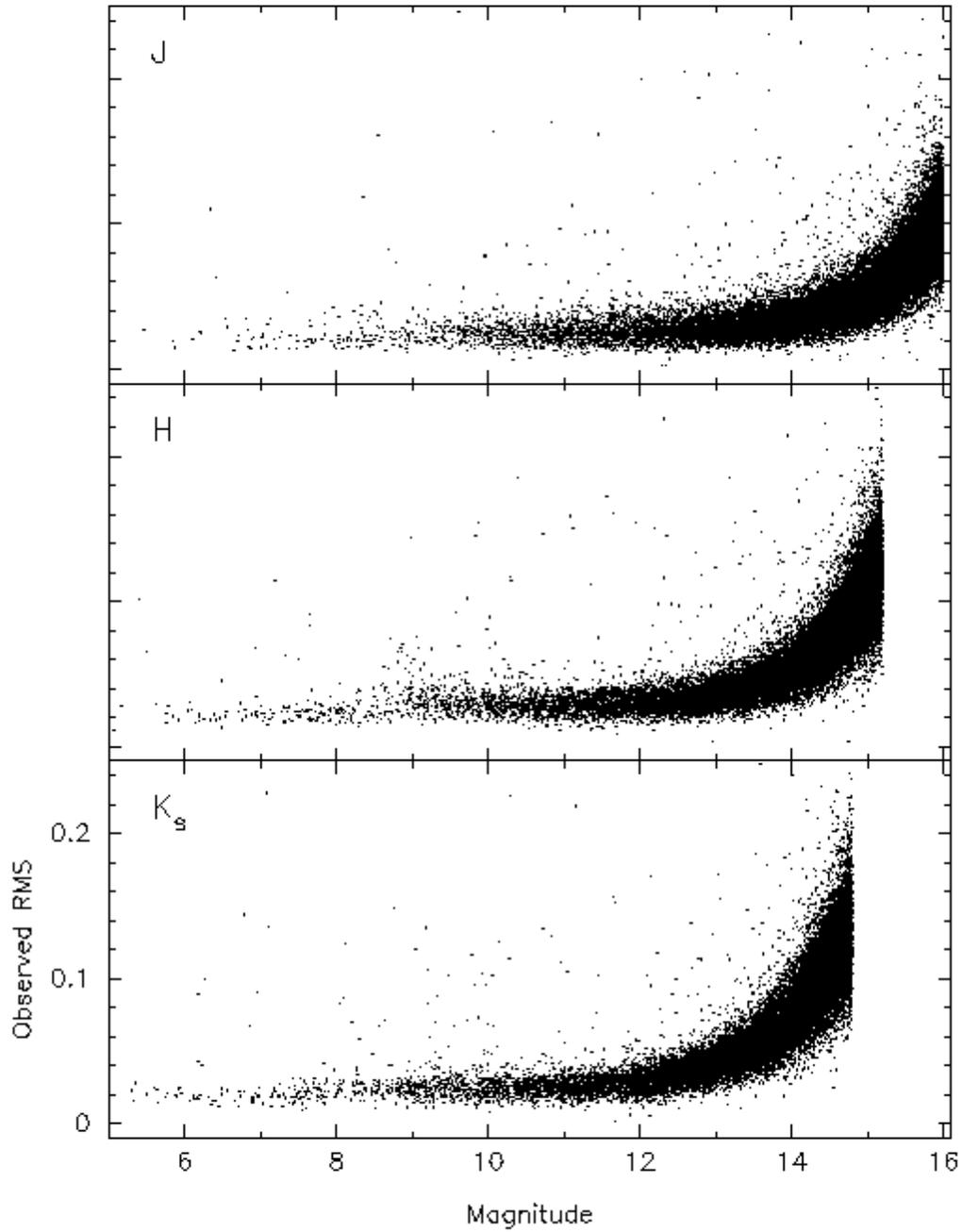}{7.3}{8.4}{0.0}{1.3}{0.85}{0}
\caption{
  The observed photometric RMS in the time series data as a function of 
  magnitude for stars brighter than the defined completeness limits. The 
  observed RMS ranges from \about 0.020 mag for the brightest stars to 
  \aboutless 0.15 mag (i.e. signal to noise ratio $\ge$ 7) for stars 
  at the completeness limit. The scale on the y-axis is the same for each
  panel.
  \label{fig:rms}
}
\end{figure}

To estimate the signal to noise of the photometry, Figure~\ref{fig:rms} shows
the observed photometric RMS in the time series for each star as a function of 
magnitude. As discussed in CHS01, suspect photometric measurements 
for stars that are extended at the 2MASS resolution or have a high reduced 
chi-squared from a single PSF measurement were excluded when computing the RMS.
Figure~\ref{fig:rms} shows a correlation with magnitude as expected if the 
observed RMS in the time series is due for the bulk of the stars to 
photometric noise and not due to intrinsic variability. The observed RMS 
values range from a minimum of \about 0.020 mag for the bright stars to 
\aboutless 0.15 mag for stars near the 
completeness limit. The observed RMS floor of \about 0.020 mag for the brighter 
stars is interpreted as the minimum photometric repeatability for these data, 
and consequently, a minimum photometric uncertainty of 0.020 mag has been 
imposed on all of the photometric measurements. Based upon the estimated 
photometric uncertainties produced by the IPAC data reduction pipeline, we 
find that 99\%, 90\%, 80\% of the stars brighter than the respective 
completeness limits at $J$, $H$, and \KB\ have a signal to noise ratio per 
measurement $\ge$ 10, and 99\% have a signal to noise ratio $\ge$ 7.

\section{Variable Stars in the Chamaeleon~I Molecular Cloud}
\label{variable_stars}

\subsection{Identification}
\label{id}

Operationally, we define as a variable any star that exhibits larger 
photometric variations over the course of a time series than expected based 
upon the photometric uncertainties. Following CHS01, we used the 
Stetson statistic \citep{Stetson96}, which correlates the photometric 
fluctuations observed in $J$, $H$, and $K_s$ bands, as our primary means to 
identify candidate variable stars. As a secondary indicator, we also 
identified stars that have large reduced chi-squared in the time series 
data but otherwise small Stetson index. The light curves and images for 
each candidate variable star were visually examined; 17 stars were removed 
and 12 stars that had a high reduced chi-squared but otherwise low Stetson index
were added. The final list contains 95 variable stars.

\begin{figure}
\insertplot{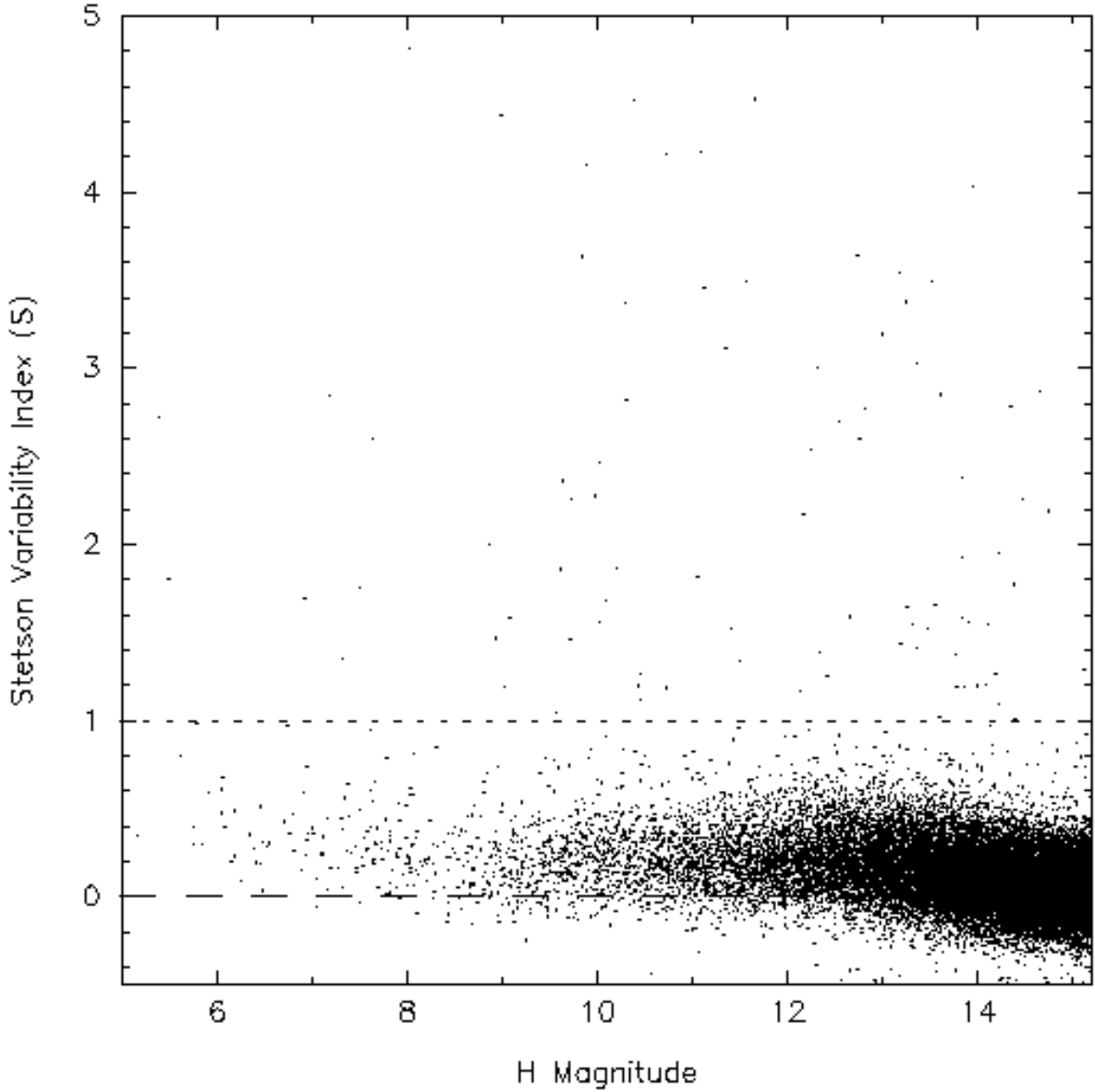}{6.8}{8.4}{0.0}{1.8}{1.0}{0}
\caption{
  The Stetson variable index ($S$) plotted as a function of the $H$ magnitude 
  for stars brighter than $H$=15.2. The dashed line at $S=0$ shows the 
  expected value of the 
  variability index for non-variable stars. The origin of the positive bias in 
  the computed index values is unknown, and suggests that a weak correlation 
  exists between the $J$, $H$, and \KB\ photometry, possibly from the fact 
  that the three bands were observed at the same time. The dotted line at 
  $S=1.00$ represents the minimum adopted value used to identify variable 
  stars in this study. Note that 7 stars with $S > 5.0$ are not shown.
  \label{fig:stetson}
}
\end{figure}

Figure~\ref{fig:stetson} shows the Stetson statistic ($S$) as a function of 
the $H$-band magnitude. For random noise, the Stetson variability index should 
be scattered around zero, and have higher, positive, values for stars with 
correlated variability. As in the Orion analysis, the Stetson variability 
index for the Chamaeleon~I objects has a positive value on average for 
brighter stars. The origin of this offset is unclear, but is suggestive of a 
weak correlation between the $J$, $H$, and \KB\ photometry, possibly due to 
the fact that the three bands are observed simultaneously at each point in the 
time series. The minimum value of the Stetson variability index which likely 
represents real photometric  variability, as opposed to random noise, was 
estimated by plotting the Stetson index versus the observed $\chi^2_\nu$, and 
also from visual examination of the light curves as a function of the Stetson 
index. Variable stars were defined based on this analysis as objects having 
Stetson index $S \ge 1.00$. This threshold is larger than the value used for 
the Orion observations ($S = 0.55$) since the Orion data contained more 
photometric observations and consequently exhibited smaller scatter in the 
Stetson index.

Table~\ref{tbl:variables} summarizes the photometric properties of the 95 
variable stars identified from our data. Included in the table are an ID 
number, a common name for previously identified variable stars, the equatorial 
J2000 coordinates, the average $J$, $H$, and \KB\ magnitudes, the observed 
photometric RMS, the number of high quality photometric measurements 
used to assess the variability, and the Stetson variability index. 
%The 
%photometric information and Stetson index reported in this table were computed 
%using all available, unflagged photometry as discussed in Section~\ref{data}.

\subsection{Light Curves}
\label{lightcurves}

To illustrate the data obtained for this study, Figure~\ref{fig:variable}
presents time-series photometry for the variable star 11344 (also known 
as T~29 and Sz~22). This figure shows the $J$, $H$, and $K$ light curves, the 
\KB\ vs. \HK\ color-magnitude diagram, and the $J-H$ vs. \HK\ color-color 
diagram. The electronic version of this article contains figures similar to
Figure~\ref{fig:variable} for all 95 variable stars, but which also include 
the $J-H$ and $H-K_s$ light curves, the $J$ vs. $J-H$ color-magnitude diagram, 
postage stamps of the $J$, $H$, and $K_s$ images, and a tabular summary of the 
photometric data. Further, {\tt .gif} images of these figures, links to 
tabular data, and cross references to existing optical and near-infrared 
catalogs are also currently available at the web site 
http://www.astro.caltech.edu/$\sim$jmc/variables/cham1. 

As with the Orion observations, the variable stars in the Chamaeleon~I
dataset show a diversity of behavior including: gradual 
increases or decreases in the stellar brightness over the course of the time 
series observations, decreases in the brightness on discrete days that may 
indicate an eclipsing system, colorless fluctuations, and instances 
where the colors become redder as the star fades. (Periodic variables were 
also identified in the Orion study but a similar time-series analysis was not 
performed on the Chamaeleon~I data due to the more limited, coarser time 
sampling.) 

Most of the Chamaeleon~I variables do not exhibit color variations with the 
brightness fluctuations. However, 14 of the variables exhibit observed 
dispersions in $J-H$ and $H-K_s$ colors more than 1.5 times larger than 
expected based on photometric noise. Figure~\ref{fig:variable} shows one 
example of color variability, which may be caused by either extinction 
variations or hot spots on the stellar surface (CHS01). After 
considering that up to half of the variable stars in Chamaeleon~I may be field 
stars (see Sections~\ref{spatial} and \ref{colors}), we conclude that \about 
15-30\% of the variables associated with Chamaeleon~I show color variability. 
Similarly, 23\% of the Orion variables show color variations of this type. 
No stars in Chamaeleon~I become bluer as they dim, which is consistent with 
the results in Orion where only 1\% of the variables exhibited these 
characteristics.

\subsection{Spatial Distribution of the Variable Stars}
\label{spatial}

The spatial distribution of variable stars identified in this study is 
presented in Figure~\ref{fig:radec}. Also shown for comparison are
(1) all sources in our point-source list with $J \le$ 16.0,
(2) sources with a near-infrared excess detectable in the $J-H$ vs 
    $H-K_s$ color-color diagram (see Section~\ref{excess}),
(3) 196 known or candidate members of the Chamaeleon~I molecular cloud 
    identified prior to this study (see Appendix),
(4) x-ray sources selected from the ROSAT all-sky survey \citep{Alcala95}, 
and 
(5) an image of the average $H-K_s$ color in the point source list binned
    to a resolution of 5$'$.
The overall density of stars increases from the south to the north, which 
reflects the decreasing distance (galactic latitude from $-18$\arcdeg\ to 
$-13$\arcdeg) to the galactic plane. Obscuration from dust in the Chamaeleon~I 
molecular cloud is clearly manifested in a sharp decrease in the $J$-band star 
counts near the center of the image and a corresponding increase in the 
average $H-K_s$ color.

Variable stars are found over the entire 6\arcdeg\ long region with a clear 
enhancement toward 2 regions in the molecular cloud at declinations of \about 
$-76.5$\arcdeg\ and \about $-77.5$\arcdeg. The declination band between 
$-76.2$\arcdeg\ and $-77.8$\arcdeg\ that encompasses these two regions 
contains 63\% (60/95) of the total number of identified variable stars despite 
containing only 27\% of the survey area. The fact that the variable star 
surface density is highest toward the molecular cloud where the overall 
stellar density is lowest suggests direct affiliation of the variable stars 
with the molecular cloud and hence a young stellar population. Indeed, 45 of 
the 95 variable stars have been previously identified as likely members of the 
Chamaeleon~I T Tauri association. Of the remaining 50 objects, 15 are 
projected against the molecular cloud and are discussed as possible 
Chamaeleon~I members in Section~\ref{candidates}, and 35 are distributed over 
a larger region outside the molecular cloud boundaries. This widespread 
variable population may represent either variable field stars, intermediate
age ($>$ 10 Myr) pre-main-sequence stars that have formed in the vicinity of 
the Chamaeleon~I molecular cloud, or young ($<$ 10~Myr) pre-main-sequence 
stars formed in Chamaeleon~I that have drifted from the molecular cloud. 
The data obtained here do not enable us to make a definitive interpretation
of these variable stars, although as discussed in the following section, the
colors and magnitudes of these stars suggest they are not likely to be a 
young stellar population related to the Chamaeleon~I molecular cloud.

\subsection{Colors and Magnitudes of the Variable Stars}
\label{colors}

The nature of the variable stars can be clarified by analyzing their colors 
and magnitudes. At the sensitivity of the 2MASS observations, the field star
population is dominated by main sequence stars of spectral type
late-F to early-K (see, e.g., Wainscoat \etal~1992). In contrast, most of
the known members of Chamaeleon~I have K and M spectral types \citep{Lawson96},
and should have redder intrinsic colors than the field stars. Moreover, given 
the proximity and youth of the Chamaeleon~I association, pre-main-sequence 
stars in Chamaeleon~I will have brighter apparent magnitudes than the typical 
field star of the same spectral type.

\begin{figure}
\insertplot{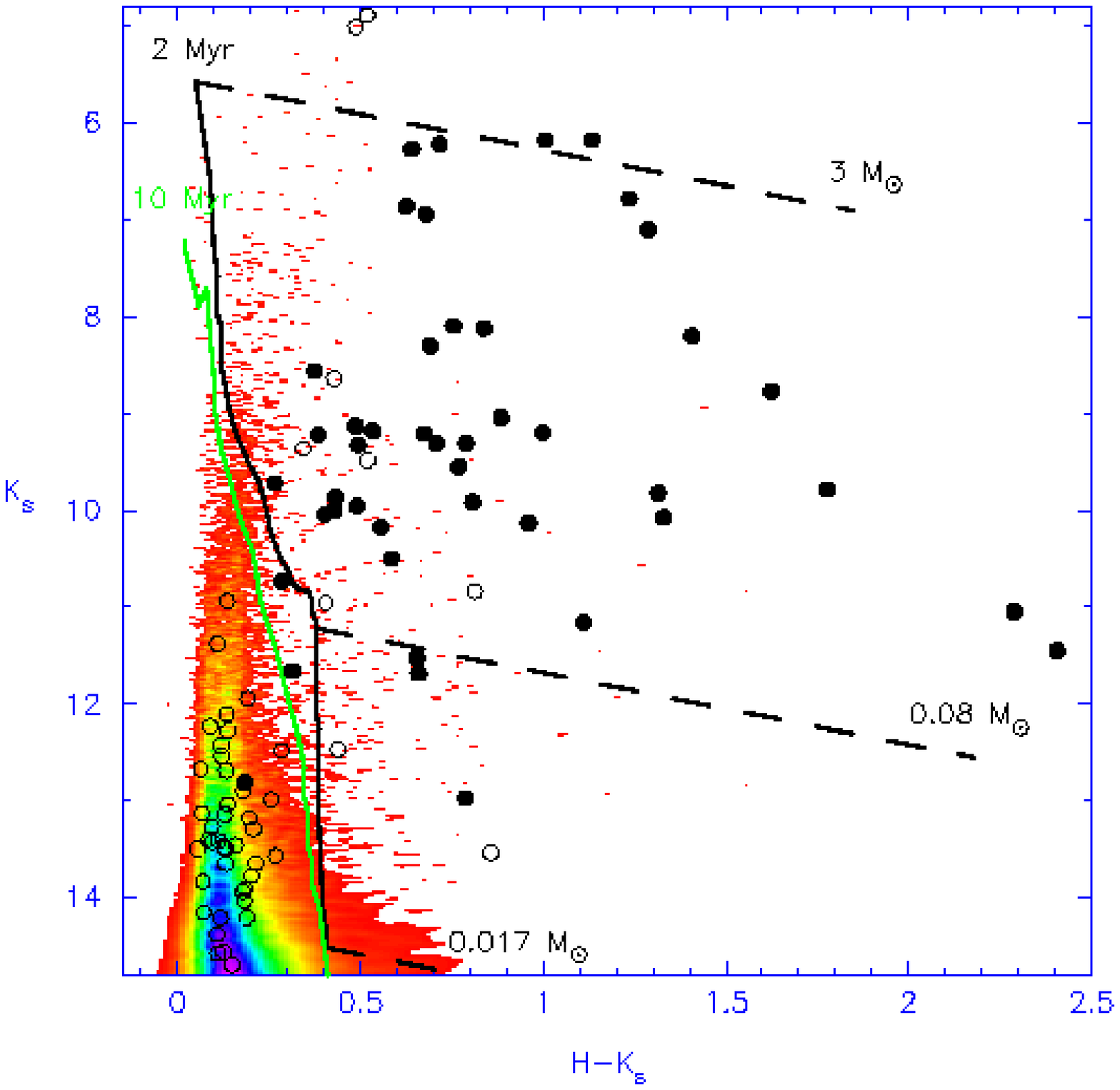}{6.8}{8.4}{0.0}{1.8}{1.0}{0}
\caption{
  \KB\ vs. \HK\ color-magnitude diagram for all stars (color scale) and 
  the variable stars (circles). Filled circles indicate stars which have
  been previously identified as members of Chamaeleon~I. The solid black curve 
  shows the 2~Myr pre-main-sequence isochrone from \citet{DM97} for masses 
  between 0.017\msun\ and 3.0\msun, and the green curve shows the 10~Myr 
  isochrone. The dashed lines indicate the reddening vector for 10 magnitudes 
  of visual extinction from \citet{Cohen81} transformed into the 2MASS 
  photometric system \citep{Carp01}, where the reddening vector is drawn
  at 0.017, 0.08, and 3.0\msun. This figure shows that a large number of the
  variable stars are consistent with reddened pre-main-sequence stars with 
  masses \aboutless 3\msun. A second group of variable stars are faint and
  blue, and as discussed in the text, are most likely variable field stars
  or old pre-main-sequence stars unrelated to Chamaeleon~I.
  \label{fig:khk}
}
\end{figure}

\begin{figure}
\insertplot{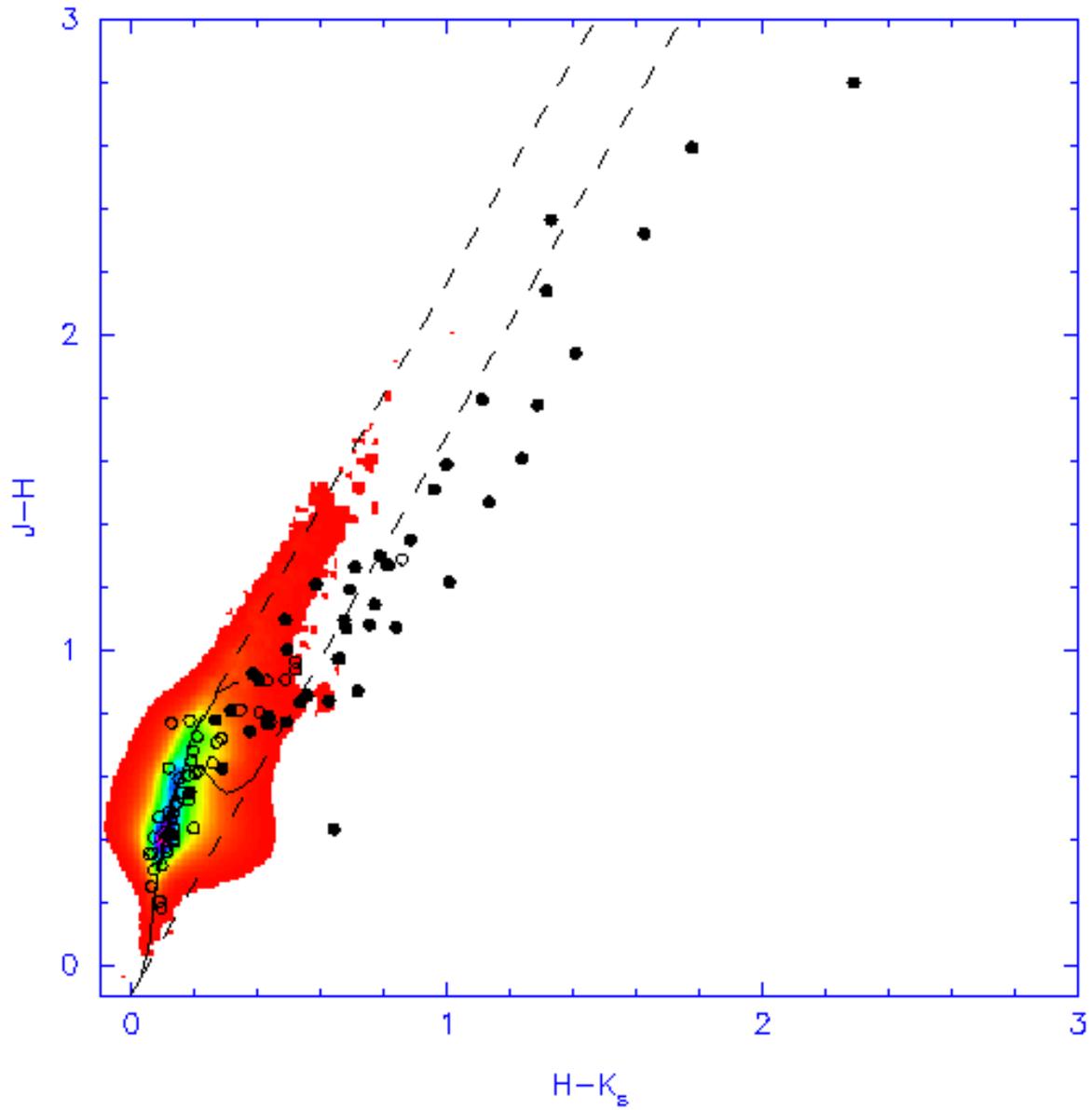}{6.8}{8.4}{0.0}{1.8}{1.0}{0}
\caption{
  $J-H$ vs. \HK\ color-color diagram for all stars (color scale) and the 
  95 variable stars (circles) identified from the $J$, $H$, and $K_s$ 
  time-series data. The filled circles represent variable stars that were 
  previously identified as likely Chamaeleon~I members.
  \label{fig:jhhk}
}
\end{figure}

To examine the colors and magnitudes for the stars detected in our 
observations, Figure~\ref{fig:khk} shows the $K_s$ vs $H-K_s$ diagram for the 
variable star population (circles) compared to all sources detected in the 
survey (represented in Hess format by the color scale). The filled circles 
indicate the subset of variable stars that have been previously identified as 
members or candidate members of Chamaeleon~I (see Appendix). The black solid 
line in Figure~\ref{fig:khk} shows the 2~Myr pre-main-sequence isochrone from 
\citet{DM97}, corresponding to the mean age of the cluster \citep{Lawson96}, 
and the dashed line shows the interstellar reddening vector. The variables can 
be coarsely grouped into two populations: 
(1) bright ($K_s$ \aboutless 12) and red ($H-K_s$ \aboutmore 0.3) stars, and 
(2) relatively faint and blue stars. Most of the bright red variables are 
known members of Chamaeleon~I with colors and magnitudes consistent with a 
young, reddened, pre-main-sequence population. Further, as shown in 
Figure ~\ref{fig:jhhk}, many of these red variables exhibit a near-infrared 
excess in the $J-H$ vs. $H-K_s$ color-color diagram consistent with the 
presence of an optically thick circumstellar accretion disk 
\citep{Lada92,Meyer97}.

\begin{figure}
\epsscale{0.35}
\plotone{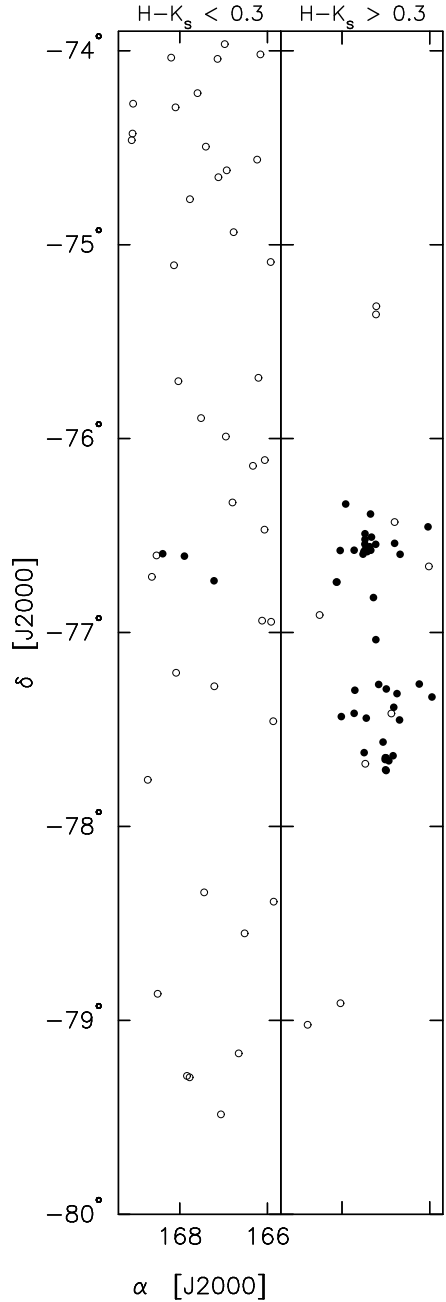}
\caption{
  Spatial distribution of variable stars for two different color ranges.
  The left panel shows the distribution of relatively blue variables with
  $H-K_s < 0.3$, and the right panel shows the distribution of red variables
  with $H-K_s > 0.3$. The blue variable stars are found over the
  entire region, while the red variable stars, not surprisingly, are located 
  mainly toward the molecular cloud. Of the 4 red variable stars found 
  well outside the cloud boundaries, one is thought to be a Mira variable,
  one is a known pre-main-sequence star identified from the ROSAT All Sky
  Survey \citep{Alcala95}, and the remaining 2 are of unknown origin. As
  discussed in the text, the faint, blue variable stars are most likely 
  field stars or an older population ($>$ 10 Myr) of low mass 
  pre-main-sequence stars located beyond the Chamaeleon~I cloud.
  \label{fig:radec_color}
}
\end{figure}

To further investigate the distinction between ``red'' and ``blue'' variable 
stars, Figure~\ref{fig:radec_color} shows the spatial distribution of variables 
separated by $H-K_s$ color, where the filled symbols again represent 
previously known Chamaeleon~I members. Not surprisingly most of the red 
variable stars are found toward the Chamaeleon~I molecular cloud and suggest 
that the red colors can be attributed to a combination of extinction, 
near-infrared excess emission, and late spectral types for many of the 
association members. Four stars with red colors are found outside the 
boundaries of the Chamaeleon~I molecular cloud. One of these variable stars is 
XZ~Cha, thought to be a Mira star. A second is a suggested optical counterpart 
to a ROSAT x-ray source RX~J1108.8-7919b, and is likely a pre-main-sequence 
star \citep{Alcala95}. A third red variable is PU~Car\footnote{\citet{Alcala95} 
identified PU~Car as RX~J1108.8-7519b. However, based on the finding charts 
from \citet{Hoff63}, PU~Car is located 2.5$'$ south of this x-ray source.},
a variable star of unknown type. The fourth red variable located off of the 
Chamaeleon~I molecular cloud is anonymous.

In contrast to the red variable stars, the blue variables are found scattered
over the entire 6\arcdeg\ declination range of the survey. Figure~\ref{fig:khk}
shows that in general these stars are too blue to be consistent with the 2~Myr
old Chamaeleon~I pre-main-sequence population. The blue variable stars may 
represent either field stars unrelated to Chamaeleon~I, or older 
pre-main-sequence stars that have formed in Chamaeleon~I or its vicinity, but 
have since dispersed around the cloud. For example, assuming a 1\kms\ velocity 
dispersion, a 10~Myr pre-main-sequence population at the distance of 
Chamaeleon~I will disperse over a 3\arcdeg\ radius region if the binding 
energy from the molecular cloud is negligible. We find near-infrared variables 
over at least 6\arcdeg\ centered on Chamaeleon~I.

To investigate whether the faint, blue variable stars could plausibly represent
an older generation of stars in the vicinity of the Chamaeleon~I molecular 
cloud, the green dotted curve in Figure~\ref{fig:khk} shows the 10~Myr 
pre-main-sequence isochrone. As stars and brown dwarfs evolve toward the 
main-sequence, they become blue and fainter, but even compared to the 10~Myr 
old isochrone, most of faint variable stars have colors that are too blue. 
While it is unlikely that the widely distribution variables are comprised of 
low mass stars $<$ 10 Myr old, their colors and magnitudes are consistent with 
an older population ($>$ 10 Myr) of low mass pre-main-sequence stars 
($< 1 M_{\odot}$) located beyond the Chamaeleon~I cloud. Such a population 
could have formed at the edge of an \ion{H}{1} spur associated with Upper Sco 
OB association, similar to the eta Cha and TW Hya associations 
\citep{Mamajek01}.

Alternatively, the blue variables have properties roughly expected from a 
main-sequence field star population in that they coincide with the peak 
density of field stars shown in Figure~\ref{fig:khk}, exhibit a north-south 
gradient in their spatial distribution (see Figs.~\ref{fig:khk} and 
\ref{fig:radec_color}), and have colors consistent with G- and K-dwarfs 
(see Fig.~\ref{fig:jhhk}). The main difficulty with this interpretation is in 
understanding the origin of the near-infrared variability if these are indeed 
old field stars. Follow-up spectroscopic observations to search for lithium in 
these sources will help distinguish if the widespread variable stars are field
objects or a $>$ 10 Myr pre-main-sequence population.

\section{Candidate Members of the Chamaeleon~I Molecular Cloud}
\label{candidates}

As summarized in the Appendix, the current census of the stellar population
in the Chamaeleon~I molecular cloud has been established based on
optical variability studies, optical spectroscopy, x-rays, and infrared 
surveys. The monitoring observations obtained here complement and expand these 
previous studies, and allow candidate members to be identified based on 
variability in the $J$, $H$, and $K_s$ time-series data and from near-infrared 
excesses in the $J-H$ vs. $H-K_s$ color-color diagram. In this section, we use 
our infrared variability data to identify new candidate members of the 
Chamaeleon~I association and comment on the completeness of the current 
stellar/substellar membership.

\subsection{Variables}

Of the 95 variable stars identified in the survey, 45 are associated with
stars that have been suggested previously as members of Chamaeleon~I. As
discussed in Section~\ref{colors}, most of the remaining 50 variable stars
are likely field stars or an intermediate age (\aboutmore 10~Myr) 
pre-main-sequence population based on their colors, magnitudes, and spatial 
distribution. However, 5 variable stars have near-infrared colors ($H-K_s > 
0.3$) and coordinates ($-77.8^\circ < \delta < -76.2^\circ$) generally 
consistent with being cluster members (see Fig.~\ref{fig:radec_color}). Of 
these 5 stars, one is YY~Cha with a magnitude of $K_s = 4.891$, and is most 
likely a Mira variable \citep{Whittet91}. The remaining four stars have not 
been previously identified as variable stars; their ID numbers, coordinates, 
and photometry are summarized in Table~\ref{tbl:candidates}. The $K_s$-band 
magnitudes range of these 4 variables range from 10.8 to 13.5, and two objects 
(9484 and 15991) have $K_s$ magnitudes and $H-K_s$ colors that imply masses 
below the hydrogen burning limit if they are 2~Myr old objects at the distance 
of Chamaeleon~I. As described in the next section, two of these variable 
stars (1715 and 15991) also exhibit an apparent near-infrared excess in the
$J-H$ vs. $H-K_s$ color-color diagram, which provides additional evidence 
that they are likely members of the Chamaeleon~I association.

\subsection{Near-infrared Excess}
\label{excess}

Previous studies have identified candidate members of Chamaeleon~I 
based on the presence of a near-infrared excess in a color-color
diagram \citep{Cambresy98,Oasa99,Gomez01}. 
The angular extent of our survey and the precision that results from averaging 
all of the photometric measurements (typically between 12 and 25 independent
observations per band per star) has produced an extensive, high 
signal-to-noise database of near-infrared photometry with which to conduct a 
more accurate census of stars with near-infrared excesses than previously
possible. For the purpose of this exercise, a star was defined to have a 
near-infrared excess if the average colors of all photometric measurements are 
such that the star is located to the right of the rightmost reddening vector 
shown in Figure~\ref{fig:jhhk}. This approach is sensitive to stars with large
enough infrared excesses to shift the observed colors outside the reddening 
vector drawn from the extrema of the color range at early O and late M 
spectral types. Stars with smaller infrared excess cannot be distinguished 
from reddened field stars in this diagram. The main-sequence locus and 
reddening vector were adopted from \cite{BB88} and \citet{Cohen81} 
respectively, and transformed into the 2MASS photometric system using the 
relations in \citet{Carp01}. The main-sequence locus from \cite{BB88} extends 
to M6 spectral types. Therefore, stars located to the right of the reddening 
vector may signify an apparent infrared excess due to photometric noise, a 
true infrared excess presumably due to an optically thick inner circumstellar 
disk, or a spectral type later than M6 but earlier than \about L9 
\citep{Davy00,Leggett02}, as the T-dwarfs have bluer infrared colors.

\begin{figure}
\insertplot{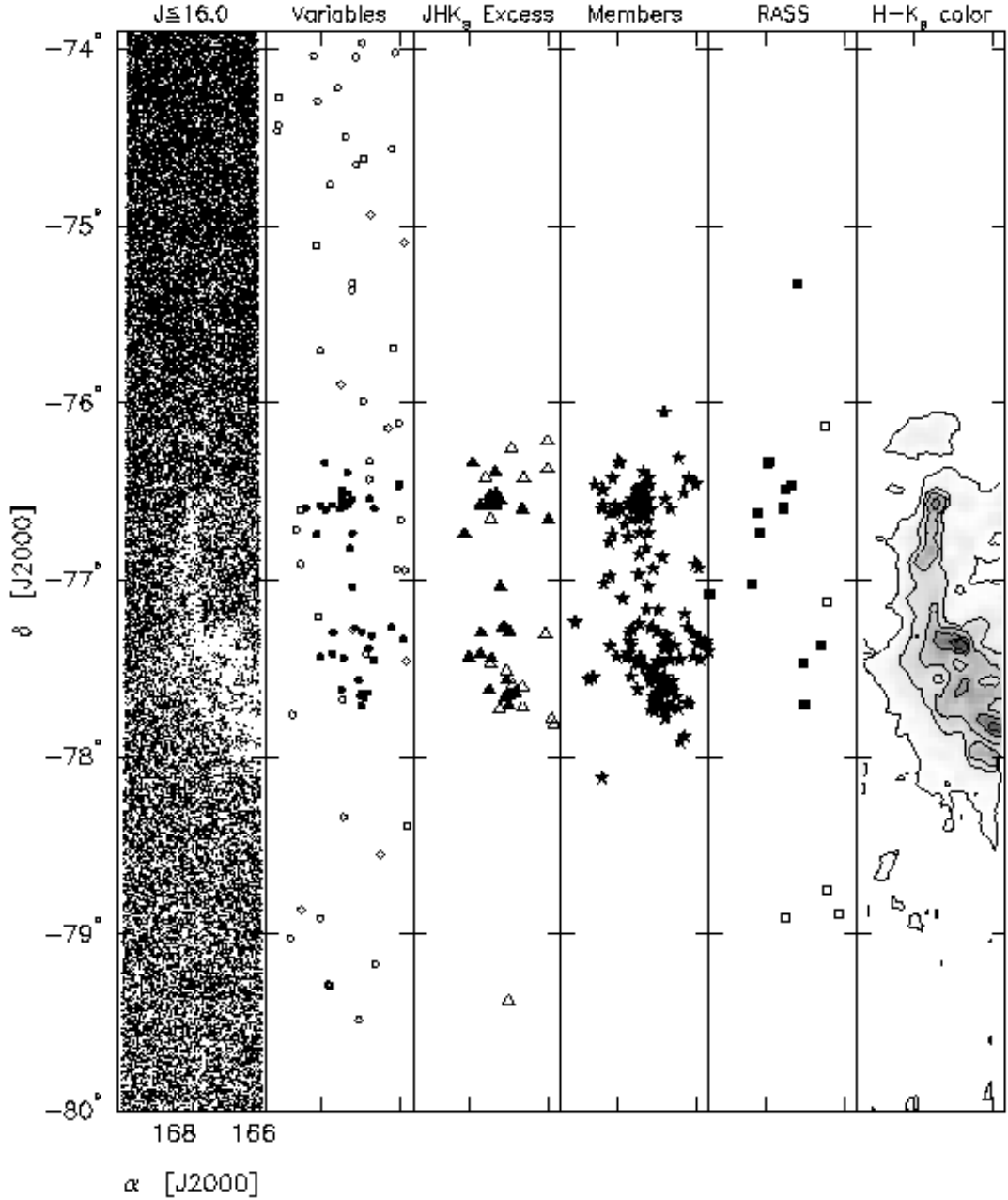}{6.6}{8.4}{0.0}{1.8}{0.9}{0}
\caption{
 Spatial distribution of stars toward the Chamaeleon~I molecular cloud.
 Starting with the leftmost panel, these figures show 
 ({\it a}) the spatial distribution of sources with $J\le 16$;
 ({\it b}) location of variable stars identified from our near-infrared data,
           where filled symbols indicate variable stars that were previously
           identified as Chamaeleon~I members;
 ({\it c}) sources that exhibit a near-infrared excess in the $J-H$ vs. $H-K_s$ 
           color-color diagram, where filled triangles represent variable 
           stars, and open triangles indicate non-variables brighter than 
           $K_s$=13.5;
 ({\it d}) members and candidate members of the Chamaeleon~I molecular cloud 
           identified prior to this study (see Appendix);
 ({\it e}) x-ray sources selected from the ROSAT All Sky Survey, where filled 
           squares represent x-ray sources that have been associated with 
           pre-main-sequence objects and open squares represent objects 
           unrelated to Chamaeleon~I \citep{Alcala95};
and
 ({\it f}) a map of the average $H-K_s$ stellar color with 5$'$ resolution, 
           where the contour levels are at 0.20, 0.35, 0.50 mag, and 
           increments of 0.30 mag thereafter.
These panels show that the largest concentration of variable stars is toward
the Chamaeleon~I molecular cloud despite the overall decrease in the stellar
surface density, indicating many of the variable stars must be associated with
the Chamaeleon~I molecular cloud. 
 \label{fig:radec}
}
\end{figure}

\begin{figure}
\insertplot{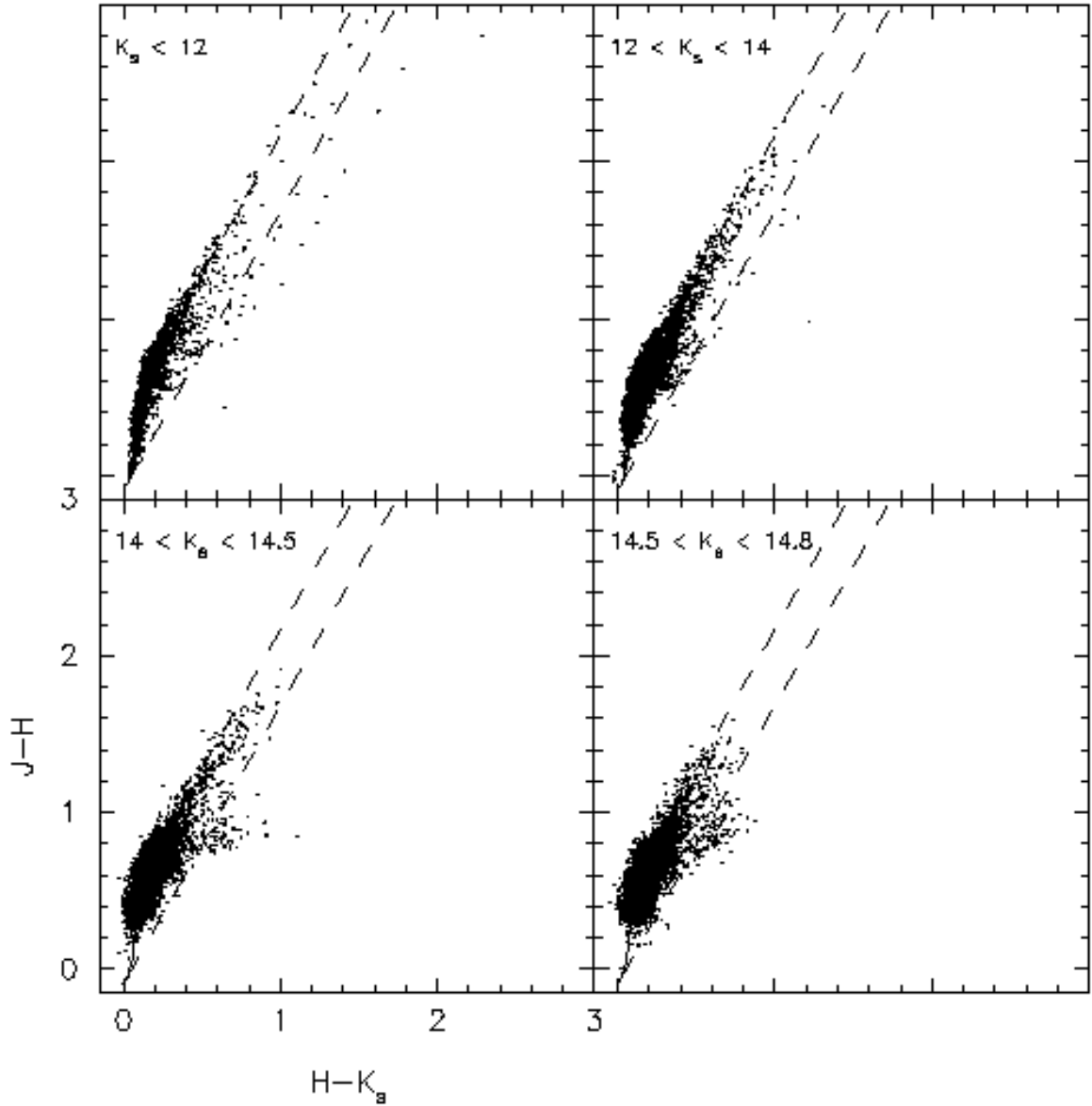}{6.8}{8.4}{0.0}{1.8}{1.0}{0}
\caption{
  $J-H$ vs. $H-K_s$ color-color diagrams in four different $K_s$-band
  magnitude intervals for objects in the source list with at least 10
  $J-H$ and $H-K_s$ measurements. The number of sources shown in each panel
  are 2869 for $K_s < 12$, 10070 for $12 < K_s < 14$, 6336 for 
  $14 < K_s < 14.5$ and 5320 for $14.5 < K_s < 14.8$. Sources with a 
  near-infrared excess and brighter than $K_s$=14 are most likely stars 
  associated with Chamaeleon~I and surrounded by an accretion disk. The 
  fainter sources with an infrared excess are most likely galaxies with
  redshifts of $z$ \aboutless 0.25 as discussed in the text.
  \label{fig:jhhk_grid}
}
\end{figure}

Figure~\ref{fig:jhhk_grid} shows $J-H$ vs. $H-K$ color-color diagrams in
4 different $K_s$-band magnitude intervals for sources with at least 10 $J-H$ 
and $H-K_s$ measurements. The later criteria was arbitrarily imposed to ensure 
high signal-to-noise photometry. Sources brighter than $K_s$=14 and with 
near-infrared excess have $J-H$ colors that range between \about 1 and 3~mag. 
Spatially these bright sources are located toward the Chamaeleon~I molecular 
cloud (see Fig.~\ref{fig:radec}). These properties are consistent with the
notion that the bright objects with infrared excesses are pre-main-sequence 
stars surrounded by optically thick accretion disks embedded within the 
Chamaeleon~I molecular cloud. 

\begin{figure}
\insertplot{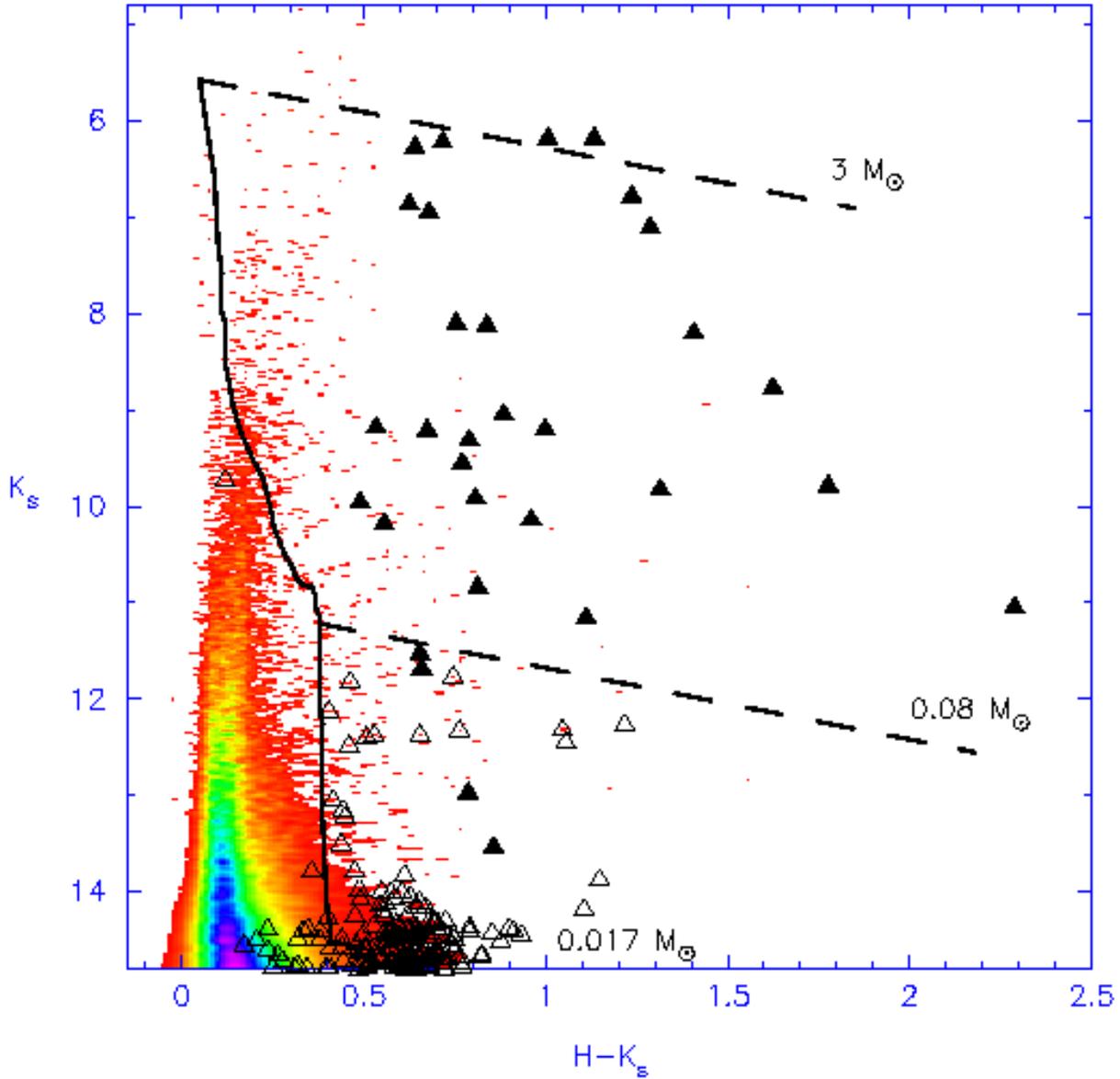}{6.8}{8.4}{0.0}{1.8}{1.0}{0}
\caption{
  $K_s$ vs. \HK\ color-magnitude diagram for all sources (color scale) and
  sources with a near-infrared excess (triangles) detectable in the $J-H$ vs. 
  $H-K_s$ color-color diagram. Filled triangles represent the subset of 
  sources with excesses that are also variable. Sources with excesses and
  brighter than $K_s$=13.5 (see Fig.~\ref{fig:radec}) are clustered around the 
  Chamaeleon~I molecular cloud and are most likely pre-main-sequence stars
  (for $K_s$ \aboutless 11) and brown dwarf candidates (for 11 \aboutless
  $K_s$ \aboutless 13.5). Objects fainter than $K_s=13.5$ are distributed
  over the entire survey area and are most likely galaxies as discussed in 
  the text.
  \label{fig:khk_excess}
}
\end{figure}

The $K_s$ vs. $H-K_s$ color-magnitude diagram for the 208 sources with an
apparent near-infrared excess is shown in Figure~\ref{fig:khk_excess},
where filled symbols denote stars with a near-infrared excess that also show 
infrared variability, and open symbols indicate non-variable sources with an
excess. Figure~\ref{fig:khk_excess} shows that all but one source with 
a near-infrared excess and brighter than $K_s$=11.5 (approximately the 
hydrogen burning limit for the distance and age of Chamaeleon~I with no 
extinction) is also variable in the near-infrared. The one star (ID number
1959) with an apparent infrared excess and is not variable has blue infrared 
colors ($J-H$,$H-K_s$ = 0.104, 0.120) and possesses a small near-infrared 
excess. The $H-K_s$ color for this star is too blue to be consistent with 
membership in the Chamaeleon~I association assuming an age of 2~Myr. We 
suspect therefore that this star most likely has an apparent near-infrared 
excess due to photometric noise, and we do not list this star as a candidate 
member of Chamaeleon~I.

Figures~\ref{fig:jhhk_grid} and ~\ref{fig:khk_excess} also show a population 
of 155 sources fainter 
than $K_s=14$ that, if stellar, exhibit a near-infrared excess. From visual 
inspection of the images, a few of these sources have suspect photometry due 
to a nearby bright star or a close companion. However, for most of the objects,
there is no a priori reason to question the photometry. The apparent excesses 
cannot be attributed to random photometric noise since one would expect that 
these objects would be present with colors ranging from $J-H$ = 0.3 to 0.9 
where the stellar density is highest in the color-color plot. Instead, most 
of the faint objects with excesses are found in a narrow range of colors 
between $J-H$ = 0.8 and 1.0. Spatially these objects are found over the full 
6$\arcdeg$ declination range of the survey. While we cannot exclude the 
possibility that these objects are very low mass objects dispersed from 
Chamaeleon~I, such a scenario would seem unlikely since it would require 
disparate spatial distributions between the substellar and the stellar 
populations, and would imply that the Chamaeleon~I IMF is heavily weighted 
toward brown dwarf objects in contrast to the IMF in other nearby star forming 
regions (see, e.g., Luhman \etal~2000). The faint infrared excess objects 
cannot be field brown dwarfs since their surface density of \about 
15~deg$^{-2}$ between $K_s=$ 14 and 14.5 is \about 300 times larger than the 
surface density of L-dwarfs down to $K_s=14.5$ \citep{Davy99}. 

The remaining possibility then is that the faint sources with near-infrared
excesses are galaxies, where the recessional velocities shift the galaxy colors 
to the right of the reddening vector in the $J-H$ vs. $H-K_s$ diagram. The 
near-infrared colors observed in Figure~\ref{fig:jhhk_grid} can be accounted 
for by a population of galaxies at redshifts between \about 0.1 and 0.25 
\citep{Mannucci01}. Photometric and spectroscopic surveys have in fact shown 
that galaxies with magnitudes of $K$=14-15 have a mean redshift of 0.18 
\citep{Son94} and a surface density of \about 100~deg$^{-2}$ \citep{V00}. 
By comparison, the observed surface density of sources in Chamaeleon~I with 
near-infrared excesses and fainter than $K_s$=14 is \about 35~deg$^{-2}$. 
The lower surface density of faint near-infrared excess sources compared to
the expected galaxy population may be attributed to the high extinction from 
the Chamaeleon~I molecular cloud and that not all of the galaxies will have
high enough redshift to produce the red near-infrared colors. We therefore 
conclude that the sources with a near-infrared excess and fainter than 
$K_s$=14 consist predominantly of galaxies with redshifts of \about 0.2.

Given the apparent predominance of galaxies among the near-infrared excess
objects at faint magnitudes, we only considered infrared excess sources 
brighter than $K_s=$13.5 as possible members of Chamaeleon~I. The spatial 
distribution of these 45 objects is shown in the third panel of 
Figure~\ref{fig:radec}. All but one is concentrated toward the cloud. The 
one exception, star 11564 in our source list (J2000 equatorial coordinates 
of $\alpha,\delta$ = 11:08:03.56,$-$79:22:34.65), has average colors of 
$J-H$=0.976 and $H-K_s$=1.218. The infrared excess appears in each of the 24 
individual photometric measurements with a PSF chi-square less than 2 in all 
observations. Therefore, the near-infrared excess cannot be merely due to 
photometric noise or a poor PSF fit. This object may either be an isolated 
young star with near-infrared excess, or a bright, red galaxy. We interpret 
the remainder of the near-infrared excess sources brighter than $K_s$=13.5 as 
young stellar objects whose excesses can be accounted for by an optically 
thick inner circumstellar disk for the brighter objects, and/or a spectral 
type later than \about M6 for the fainter objects. Table~\ref{tbl:candidates} 
lists the seven new near-infrared candidates members identified from our 
observations that are located in the vicinity of the Chamaeleon~I molecular
cloud. (The eight star with an infrared excess in Table~\ref{tbl:candidates} 
is fainter than our imposed magnitude limit, but is included in the table as 
a candidate members because it is a variable.) 

We note that we did not confirm many of the near-infrared excess stars 
identified in previous studies. \citet{Cambresy98} selected 54 candidate 
members in Chamaeleon~I based on near-infrared excess in the $I-J$ vs. 
$J-K$ diagram or redder colors than expected from a Chamaeleon~I extinction 
map. Of these 54 stars, 42 are in our database with three-band photometry, 
and only 6 show an infrared excess in our data. One of these 
sources is identified in Table~\ref{tbl:candidates}, and three others 
(ID numbers 32, 41, and 49 in Table~2 of Cambresy \etal~1998) are indicated as 
candidate members in the Appendix since they have other characteristics that 
suggest they may be pre-main-sequence stars. The remaining two sources (ID 
numbers 329 and 926 in our study, and ID 10 and 11 respectively in Cambresy 
\etal~1998) show infrared excess in our data as well, but they were detected 
on only 5-6 $J$-band images due to the faintness of the object and were not 
included in the near-infrared excess analysis. Similarly, \citet{Gomez01}
list 56 infrared excess candidates (identified from the $J-H$ vs. $H-K_s$ 
diagram) brighter than $K_s$=14, of which 49 are three band detections in our
observations. Only 7 of these sources show an infrared excess. One of these
objects is indicated in Table~\ref{tbl:candidates}, four others (ID numbers
13, 29, 30, and 31 in Gomez \& Kenyon~2001) are listed in the Appendix as 
members based on other studies, and 2 (ID numbers 10 and 40 in Gomez \& 
Kenyon~2001) are fainter than $K_s=$14 in our study and did not meet our 
magnitude criterion. (As shown in CHS01, some stars do show a transient 
near-infrared excess. However, it is unlikely that this account can account 
for the majority of the Gomez \& Kenyon sources since an equally large number 
of new near-infrared excesses should have been identified in our observations, 
which is not observed.) Finally, \citet{Oasa99} list 19 sources with infrared
excesses, of which 12 are in our database. The remaining 7 sources are fainter
than our sensitivity limits. Of these 12 sources, 7 show an infrared excess.
All of the infrared excess candidates from \citet{Oasa99} are listed in the 
Appendix.

\subsection{Discussion}

The area covered by our observations encompasses 170 known or candidate 
Chamaeleon~I members identified prior to this study, of which 159 have been 
detected in the near-infrared with our data. As discussed in CHS01, 
the monitoring observations discussed here
are most sensitive to detecting variables brighter 
than $K_s$ \about 12. Of the 129 Chamaeleon~I members brighter than this 
limit, 43, or 33\%, have been detected as variable. The variables include stars
with and without near-infrared excesses as traced in the $J-H$ vs. $H-K_s$ 
diagram, although the variability observations are biased toward detecting 
stars with excesses. That is, of the Chamaeleon~I members brighter than 
$K_s$=12, 23\% (30/129) have a near-infrared excess, but 28/43 (65\%) of the 
variables have an excess. Thus the proportion of bright stars that have a 
near-infrared excess is higher among the variable compared to the Chamaeleon~I
members at a \about 5$\sigma$ confidence level. Similarly, in Orion, 
19\%$\pm$1\% of the stars brighter than $K_s$=12 exhibit a near-infrared 
excess, compared to 28\%$\pm$2\% of the variables brighter than this limit. 
While the Orion variables are also slightly biased toward stars with infrared
excesses (at the \about 4.5$\sigma$ confidence level), the bias is not as 
strong in that the overall percentage of variables with an infrared excess is 
lower in Orion by a factor of 2.3$\pm$0.3. 
% Orion statistics are as follows:
% 277/1489 stars with K<=12 have a near-ir excess
% 166/599 variables with K<=12 and in main cluster have near-ir excess

The fact that nearly all Chamaeleon~I members with a near-infrared excess and
brighter than $K_s$=12 are also variable suggests that the variability may be
related to the presence of an inner accretion disk. Further evidence for this
conjecture comes from the type of variability exhibited by stars with infrared
excesses. Of the 14 stars in Chamaeleon~I that show measurable color 
variability where the stellar color becomes redder as the star fades (see 
Section~\ref{lightcurves}), 12 have a near-infrared excess. These 
near-infrared color variations cannot be readily accounted for by cool spots 
on the stellar surface (Skrutskie \etal~1996; CHS01). They can however, be 
explained by extinction variations (perhaps from a warped circumstellar disk),
or hot spots on the stellar surface produced by accretion from the disk onto 
the star. However, hot spots can only produce colors variations of up to
\about 0.1 mag in the near-infrared given typical hot spot parameters modeled 
in T Tauri stars, and extinction variations may be the preferred explanation
for the large amplitude variables (see CHS01). More extensive observations and 
modeling of these color variations may provide a unique window into the 
properties of the inner circumstellar disk.

Two of the Chamaeleon~I variables (ID numbers 18416 and 21473) that show color 
variations have masses near the hydrogen burning limit as inferred from 
the $K_s$ vs. $H-K_s$ diagram. This suggests that the variability mechanisms
discussed in context of the Orion observations for stars more massive than
1\msun\ may extend to lower mass objects. For fainter sources between $K_s$ 
magnitudes of 12.0 to 13.5, and presumably lower mass, only 8\% (1/13) of the 
near-infrared excess objects also show variability. The lack of variability
in these sources does not necessarily imply a change in the infrared 
excess-variability relation for substellar objects however. Not only are these
observations less sensitive to variability for sources fainter than $K_s$=12,
but the apparent infrared excess may be due to a spectral type later than M6
and not to a circumstellar disk.

Finally, we briefly discuss the implication of these observations for the
current census of the stellar and substellar population in Chamaeleon~I. 
The region we surveyed contains 129 known or candidate Chamaeleon~I members 
brighter than $K_s$=12. Only 2 new candidate members brighter than $K_s=12$ 
have been identified in this study from either variability or a near-infrared 
excess. Since nearly all bright objects with a near-infrared excess are also
variable, these observations suggest that the current census of Chamaeleon~I
sources is nearly complete for objects brighter than $K_s$=12 and with an
infrared excess in the $J-H$ vs. $H-K_s$ diagram. 
\citet{Comeron00} recently identified 13 low mass stars and brown dwarfs in 
Chamaeleon~I from an H$\alpha$ prism survey. Nine of these sources have 
magnitudes of $10.6 < K_s < 12$, but none were detected as variable, and only 
one was identified as distinctive in this dataset because of an apparent 
infrared excess. The other 4 sources in \citet{Comeron00} are fainter than
$K_s$=12, and interestingly, 3 were identified here as having a near-infrared 
excess, where the excess for these objects may be due to a late spectral type.
While the statistics are small, the comparison with \citet{Comeron00} suggests
that the $J-H$ vs. $H-K_s$ diagram is fairly efficient at identifying the 
substellar objects. The fact that only 8 new candidate members were identified 
with magnitudes of $12 < K_s < 13.5$ suggests brown dwarfs will not 
substantially expand the current population census of Chamaeleon~I.

\section{Summary}
\label{summary}

We have conducted a $J$, $H$, and $K_s$ variability study of stars in a 
$0.72^\circ\times6^\circ$ region centered on the Chamaeleon~I molecular cloud 
and T-association. Observations were obtained on 2 nights in January 2000 and 
13 nights in April and May 2000 using the 2MASS South telescope. Compared to 
our variability study of the Orion~A molecular cloud (CHS01), which was 
sensitive to variable stars more massive than \about 1\msun, observations 
of the closer Chamaeleon~I star forming region permit variability to
be detected in lower mass stars and brown dwarfs down to \about 0.05\msun.

Of the 34,539 sources meeting the photometric completeness criteria, 95 
have been identified as variable from either a large \citet{Stetson96} 
variability index or a large reduced chi-square in the time series 
data. The variables can be coarsely grouped into a population of 
bright ($K_s < 12$), red ($H-K_s > 0.3$) stars and a population of faint 
($K_s < 13$), blue ($H-K_s < 0.3$) stars. Most of the ``red'' variables are
known members of the Chamaeleon~I association, and as expected have 
near-infrared colors and magnitudes consistent with a young, pre-main-sequence
population. The ``blue'' variables are distributed over the full 6\arcdeg\ of 
the survey area. The colors and magnitudes of the blue variables are 
inconsistent with a pre-main-sequence population as old as 10~Myr at the 
distance of Chamaeleon~I, but can be explained as a population of older
pre-main-sequence stars unrelated to Chamaeleon~I or variable field stars.

The time-series data were used to identify new candidate members of the 
Chamaeleon~I association that show photometric variability and/or a 
near-infrared excess characteristics of an optically thick circumstellar disk.
Of the 95 variables identified in this study, 45 were known prior to 
this study as members or candidate members of Chamaeleon~I. Among the 
remaining 50 variables, we have identified 4 new sources that have colors and 
coordinates consistent with low mass, pre-main-sequence members of the 
Chamaeleon~I association. A total of 208 sources brighter than $K_s$=14.8 were 
identified as having a near-infrared excess in the $J-H$ vs. $H-K_s$ 
color-color diagram. The fainter sources with near-infrared excesses ($K_s < 
14$) are randomly distributed over the 6\arcdeg\ survey region. Based on their 
spatial distribution, location in the color-color diagram, and surface 
density, we suggest that these objects are most likely galaxies with redshifts 
of $z$ \aboutless 0.25. The 45 sources brighter than $K_s$=14.5 and with 
infrared 
excesses are clustered spatially around the Chamaeleon~I molecular cloud and 
are likely association members. Seven of these relatively bright sources with 
excesses are new candidate members of Chamaeleon~I, including one which is 
also variable. In total, 10 new candidate members have been identified, 8 of 
which have colors and magnitudes consistent with young brown dwarfs at the 
distance of Chamaeleon~I. These observations suggest that the current census
of the Chamaeleon~I population is complete for objects brighter than $K_s$=12
and with a near-infrared excess in the $J-H$ vs. $H-K_s$ diagram.

As a product of this study, these observations have yielded a precise dataset
of photometry and astrometry for previously identified Chamaeleon~I members.
In the Appendix we summarize the $J$, $H$, and $K_s$ magnitudes and 
cross-identifications for Chamaeleon~I members.

\acknowledgements

We would like to thank the 2MASS Observatory Staff and the Data Management 
Team for acquiring and pipeline processing the special survey observations 
used in this investigation. This publication makes use of data products from 
the Two Micron All Sky Survey, which is a joint project of the University of 
Massachusetts and the Infrared Processing and Analysis Center, funded by the 
National Aeronautics and Space Administration and the National Science 
Foundation. 2MASS science data and information services were provided by the 
InfraRed Science Archive (IRSA) at IPAC. This research has made use of the 
SIMBAD database, operated at CDS, Strasbourg, France. JMC acknowledges support 
from Long Term Space Astrophysics Grant NAG5-8217 and the Owens Valley Radio 
Observatory, which is supported by the National Science Foundation through 
grant AST-9981546.

\clearpage

\appendix

\section{Astrometry and Photometry for Previously Identified Members of 
Chamaeleon~I}
\label{members}

During the course of this study, a membership list for Chamaeleon~I was 
compiled using published observations. This Appendix presents the compilation
of known and candidate members of the Chamaeleon~I association, and summarizes
the astrometry and $J$, $H$, and $K_s$ photometry obtained from our 
observations.

The Chamaeleon~I association was initially identified as a spatial 
concentration of optical variable stars \citep{Hoff62}. Subsequent studies 
expanded the association membership through objective prism or CCD H$\alpha$ 
emission-line surveys \citep{HM73,Schwartz77,Hartigan93} and spectroscopic 
follow-up of individual stars \citep{Appenzeller77,Rydgren80,Appenzeller83} to 
identify stars with spectral features characteristic of pre-main-sequence 
objects. More recent objective prism surveys, followed by spectroscopic 
observations and multi-wavelength imaging, have begun to probe the Chamaeleon~I 
population in the substellar regime \citep{Comeron99,Comeron00}. Additional 
pre-main-sequence candidates have been identified in x-rays from Einstein 
\citep{Feigelson89} and ROSAT \citep{Feigelson93,Alcala95}, where 
pre-main-sequence counterparts to the x-ray sources have been identified from 
optical spectroscopic follow-up observations 
\citep{Walter92,HLF94,Alcala95,Lawson96}. Candidate cloud members that remain 
deeply embedded in the cloud have been identified from the presence of excess
near-infrared emission \citep{Hyland82,Jones85,Cambresy98,Oasa99,Gomez01}, red 
far-infrared colors from IRAS 
\citep{Baud84,Assendorp90,Prusti91,Whittet91,Gauvin92}, and more recently, red 
mid-infrared colors from ISO \citep{Nordh96,Persi00,Lehtinen01}. In total, the 
candidate members identified by infrared excesses can more than double the 
known stellar population.

Table~\ref{tbl:members} summarizes the status of the stellar and substellar 
membership in the Chamaeleon~I molecular cloud prior to this study. We include 
in Table~\ref{tbl:members} stars that are optically variable or show H$\alpha$ 
emission as summarized by \citet{Whittet87} and references therein for 
pre-1987 observations, the H$\alpha$ objects and red stars from 
\citet{Hartigan93}, the brown dwarfs from \citet{Comeron00}, candidate members 
identified from IRAS \citep{Baud84,Whittet91} and ISO \citep{Persi00}, and 
optical counterparts to x-ray sources that have spectroscopic signatures of 
pre-main-sequence stars \citep{Walter92,Lawson96}. Table~\ref{tbl:members} 
also included possible deeply embedded members of Chamaeleon~I that have 
recently been identified by the presence of excessed near-infrared 
\citep{Oasa99} and mid-infrared emission \citep{Persi00}. \citet{Cambresy98} 
and \citet{Gomez01} have also identified \about 150 candidate members based on 
the presence of near-infrared excess emission (and, in the case Cambresy
\etal~1998, red sources without necessarily an infrared excess). Since the
majority of the near-infrared excess candidates could not be confirmed with 
our data (see Section~\ref{excess}), these objects are not listed in 
Table~\ref{tbl:members} unless that stars contains other characteristics 
indicating it is a pre-main-sequence object (see also 
Table~\ref{tbl:candidates}).

Column~1 in Table~\ref{tbl:members} lists the commonly adopted name of the
Chamaeleon~I members. Due to the multitude of observations of the Chamaeleon~I 
region, many of the sources have been observed as part of several studies,
and the cross identifications of the various names are provided in
Table~\ref{tbl:crossids}. Most of the cross identifications were made by
matching the coordinates in the original references. Two notable exceptions
are that the positions for sources in the General Catalog of Variable stars
were taken from \citet{Lopez90} when possible, and the association of IRAS
sources from \citet{Baud84} were taken from \citet{Lawson96}. Columns 2 and 3 
list the J2000 equatorial coordinates, where the coordinates have been 
obtained from, in order of preference,
(1) this study, 
(2) the 2MASS Working Database (Version 2 processing) for stars outside our 
survey coverage, and 
(3) the literature. 
The coordinates listed in \citet{Oasa99} were found to differ from 2MASS by 
up to 10\arcsec. For sources detected only by \citet{Oasa99}, an astrometric
correction was applied based on nearby sources detected both by our observations
and \citet{Oasa99}. Columns 4-12 list the mean $J$, $H$, and $K_s$ magnitudes,
photometric RMS, and number of measurements. Column~13 lists the Stetson 
variability index derived from our data. 2MASS photometry for Chamaeleon~I
members located outside our survey boundaries have been obtained from the
Version 2 2MASS Working Database. These sources can be identified in 
Table~\ref{tbl:members} as having no Stetson index and with N$\le$=1 in each 
band. The RMS in such instances represent the photometric uncertainty from the 
2MASS data reduction pipeline excluding zero point calibration uncertainties, 
which are typically \about 1\% in each band. 

As noted by \citet{Schwartz91}, many of the sources are only suspected members 
of the cloud and have yet to be verified spectroscopically to be 
pre-main-sequence stars. In particular, some stars have been identified as 
Chamaeleon~I members because they are variable or have red photometric 
colors (but without an infrared excess), but no additional evidence has been 
obtained to indicate these are not simply field stars. Accordingly, the
flags in Column~11 indicate the properties associated with each source that
suggests it is a pre-main-sequence object. The flags indicate, in order from 
left to right,
(1) known variables from the General Catalog of Variable Stars or this study;
(2) H$\alpha$ emitting objects from \citet{HM73}, \citet{Schwartz77};
    \citet{Hartigan93}, \citet{Walter92}, \citet{HLF94}, \citet{Lawson96},
    \citet{Comeron99}, and \citet{Comeron00}, and objects with Pa$\beta$ or 
    Br$\gamma$ in emission from \citet{Gomez02};
(3) stars with lithium in absorption from \citet{Walter92}, \citet{HLF94},
    and \citet{Lawson96};
(4) x-ray sources from \citet{Feigelson89} and \citet{Feigelson93};
(5) near- or mid-infrared excesses from \citet{Oasa99}, \citet{Persi00}, and
    this study;
and
(6) far-infrared sources from \citet{Baud84}, \citet{Assendorp90}, 
    \citet{Prusti91}, and {\it IRAS} Point Source Catalog.
The flag 'Y' indicates the characteristic has been identified with that star
in at least one study; The flag `?' indicates that the authors of the original
paper expressed uncertainty of the detection; and the flag '0' indicates the
property has not yet been identified.

{\it Notes on individual sources:}

{\it UX Cha} --- \citet{Feigelson89} associated UX~Cha (= T~22) with the x-ray
source CHX~8. However, as originally noted by \cite{Feigelson93}, UX~Cha
is actually located 26$''$ to the north. The finder chart in 
\citet{Schwartz91} identifies the incorrect star as UX~Cha.

{\it C9-1, C9-2, and C9-3} --- \citet{Hyland82} identified these objects
as Chamaeleon~I members based on bolometer data. Each of these objects is
located in the bright Infrared Nebula (IRN). However, none of these
sources are visible on the 2MASS images. While it is possible that these
objects are variable and have faded from view, we find it more likely that
they are knots in nebulosity and have not included these objects in the
membership list.

{\it T 2} --- This source has been traditionally identified as a member of
Chamaeleon~I based on its optical variability. However, \citet{WB96} 
showed that this source is a RR Lyra star unrelated to Chamaeleon~I. 
This source is not listed in Table~\ref{tbl:members}.

{\it T 36} --- This source has traditionally been associated with
Chamaeleon~I based on its optical variability, but the near-infrared colors 
and magnitudes of this star ($K_s$=12.809, $H-K_s$=0.184 are inconsistent
with membership. While this star is listed in Table~\ref{tbl:members}, 
spectroscopic observations are needed to ascertain membership.

{\it T49, CHX~18N, and IRAS 11101-7603} --- The IRAS source is closest to 
CHX~18N, but within the IRAS astrometric uncertainties may also be associated 
with T~49. It is listed under both objects in Table~\ref{tbl:members}.

{\it T 52 and T 53} --- \citet{Feigelson89} lists both sources as possible 
optical counterparts to the x-ray source CHX~19. Subsequent high resolution
observations with ROSAT show the x-ray emission originates from T~52.
\citep{Feigelson93}. Therefore, we have not listed T~53 as a counterpart
to CHX~19 in Table~\ref{tbl:crossids}.

%{\it Hn12W, Hn12E, and ISO~244} --- ISO 244 is closer to Hn12E than Hn12W, 
%although it is associated with Hn12W here since it is much brighter of the
%two objects.

{\it CHXR~30 and B~38} --- \citet{Lawson96} associates CHXR 30 with B~38,
while \citet{Cambresy98} and \citet{Kenyon01} list CHXR~30 and B~38 as 
separate sources. Two stars separated by 9.9\arcsec\ are present on the 2MASS 
images.
The eastern source is consistently identified as CHXR~30 in these studies. It 
is the brighter of the two 2MASS sources in the optical with a $J-K_s$ color of 
2.70. This star has not been conclusively been identified as pre-main-sequence
object through spectroscopy, although \citet{Persi00} identified a mid-infrared
excess in this source from ISO observations. The western source, identified as 
B38 \citep{Baud84} by \citet{Cambresy98} and \citet{Kenyon01}, is the redder 
of the two 2MASS objects with $J-K_s$=3.45. This source is detected as a 
variable with a near-infrared excess in our observations. Since the beam of 
the IRAS observations used by \citet{Baud84} encompasses both 2MASS sources, 
and the position uncertainty of the ROSAT x-ray observations is 8$\arcsec$, it 
is not clear which, or if both, sources contribute to the far-infrared and 
x-ray emission. In Table~\ref{tbl:members}, we list the eastern source as 
CHXR~30a and B~38a, and the western source as CHXR30b and B~38b.

{\it HM~8 and ISO 10} --- \citet{Persi00} tentatively matched ISO~10 with 
HM~8, but they noted the larger than average positional differences between
the two objects. The positional difference is larger than the tolerance 
limit used here to match the ISO catalog with the 2MASS astrometry, and this 
association was not made in Table~\ref{tbl:members}.

%\documentstyle[aj_pt4]{article}
%
%\begin{document}

\newcommand{\h}{$^{\rm h}$}
\newcommand{\m}{$^{\rm m}$}
\newcommand{\s}{$^{\rm s}$}

%\tablenum{1}
%\pagestyle{empty}

\begin{deluxetable}{cccc}
\tablewidth{275pt}
\tablecaption{Coordinates of Observed Tiles\label{tbl:coords}}
\tablehead{
  \colhead{Tile}                       & 
  \colhead{Right Ascension\tablenotemark{a}} &
  \multicolumn{2}{c}{Number of Nights Observed} \\
  \cline{3-4} \\
  \colhead{}         & 
  \colhead{(J2000)}  &
  \colhead{Complete} &
  \colhead{Partial}
}
\startdata
1 &  11\h04\m38\s & 18 & 3\\
2 &  11\h06\m15\s & 22 & 3\\
3 &  11\h07\m46\s & 26 & 3\\
4 &  11\h09\m23\s & 28 & 3\\
5 &  11\h10\m55\s & 26 & 3\\
6 &  11\h12\m31\s & 21 & 3\\
7 &  11\h14\m02\s & 16 & 3\\
8 &  11\h15\m39\s & 16 & 3\\
\enddata
\tablenotetext{a}{Each tile is $8.5'\times6$\arcdeg\ in size, and is centered on
a declination of $-77$\arcdeg\ and the right ascension listed.}
\end{deluxetable}
\clearpage
%\end{document}

%\documentstyle[preprint]{aastex}

%\begin{document}

%\tablenum{2}
%\pagestyle{empty}

\begin{deluxetable}{rccccccccl}
\tabletypesize{\scriptsize}
\tablewidth{420pt}
\tablecaption{Observing Log\label{tbl:log}}
\tablehead{
\colhead{UT Date}  & 
\colhead{Tile 8}  & 
\colhead{Tile 7}  & 
\colhead{Tile 6}  & 
\colhead{Tile 5}  & 
\colhead{Tile 4}  & 
\colhead{Tile 3}  & 
\colhead{Tile 2}  & 
\colhead{Tile 1}  &
\multicolumn{1}{l}{Comments}
}
\startdata
January 13, 2000 & x & x & x & x & x & x & x & x & $\delta < -78^\circ$ only\\
January 25, 2000 & x & x & x & x & x & x & x & x & $\delta > -78^\circ$ only\\
April 8, 2000    & x &   & x & x & x & x & x & x &\\
April 12, 2000   & x & x & x & x & x & x & x & x &\\
April 19, 2000   &   &   &   &   & x & x & x & x &\\
April 21, 2000   & x & x & x & x & x & x & x & x &\\
April 22, 2000   & x & x & x & x & x & x & x & x &\\
April 23, 2000   & x & x & x & x & x & x & x & x &\\
April 24, 2000   & x & x & x & x & x & x & x & x &\\
April 25, 2000   & x & x & x & x & x & x & x & x &\\
April 27, 2000   & x & x & x &   & x &   &   &   &\\
April 29, 2000   & x & x & x & x & x & x & x & x &\\
May 1, 2000      & x & x & x & x & x & x & x & x &\\
May 12, 2000     & x & x & x & x &   &   &   &   &\\
May 19, 2000     & x & x & x & x & x & x & x & x &\\
\enddata
\end{deluxetable}
\clearpage

%\end{document}

%\documentclass[preprint]{aastex}
%\begin{document}

%\tablenum{3}
%\pagestyle{empty}

%\rotate

\begin{deluxetable}{rccc@{\extracolsep{10pt}}
 r@{\extracolsep{8pt}}r@{\extracolsep{8pt}}r@{\extracolsep{10pt}}
 c@{\extracolsep{8pt}}c@{\extracolsep{8pt}}c@{\extracolsep{8pt}}
 c@{\extracolsep{0pt}}c@{\extracolsep{0pt}}c@{\extracolsep{0pt}}
 @{\extracolsep{0pt}}c}
\tabletypesize{\scriptsize}
\tablewidth{530pt}
\tablecaption{Near-Infrared Variable Stars\label{tbl:variables}}
\tablehead{
  \colhead{ID}        & 
  \colhead{Name}      & 
  \multicolumn{2}{c}{Coordinates (J2000)} &
  \multicolumn{3}{c}{Mean Magnitudes}     &
  \multicolumn{3}{c}{Observed RMS}        &
  \multicolumn{3}{c}{N}                   &
  \colhead{Variability}                 \\
  \cline{3-4}
  \cline{5-7}
  \cline{8-10}
  \cline{11-13}
  \colhead{}  &
  \colhead{}  &
  \colhead{$\alpha$}  &
  \colhead{$\delta$}  &
  \colhead{$J$}       &
  \colhead{$H$}       &
  \colhead{$K_s$}     &
  \colhead{$J$}       &
  \colhead{$H$}       &
  \colhead{$K_s$}     &
  \colhead{$J$}       &
  \colhead{$H$}       &
  \colhead{$K_s$}     &
  \colhead{Index}
}
\startdata
   198 & \multicolumn{1}{c}{...} &  165.860056 & $ -78.387932$ & 13.805 & 13.361 & 13.174 &  0.140 &  0.131 &  0.113 &  12 &  12 &  12 & $  3.03$\\
   241 & \multicolumn{1}{c}{...} &  165.868449 & $ -77.457493$ & 13.889 & 13.254 & 12.988 &  0.094 &  0.080 &  0.076 &  13 &  13 &  13 & $  1.64$\\
   670 & \multicolumn{1}{c}{...} &  165.915725 & $ -76.944871$ & 14.014 & 13.605 & 13.489 &  0.131 &  0.148 &  0.134 &  13 &  13 &  13 & $  2.85$\\
   776 & \multicolumn{1}{c}{...} &  165.925563 & $ -75.089402$ & 13.678 & 13.078 & 12.898 &  0.053 &  0.053 &  0.060 &  13 &  13 &  13 & $  1.27$\\
  1012 & Hn 2                 &  165.948566 & $ -77.332334$ & 11.353 & 10.449 & 10.044 &  0.044 &  0.051 &  0.048 &  13 &  13 &  13 & $  1.27$\\
  1715 & \multicolumn{1}{c}{...} &  166.017772 & $ -76.659153$ & 12.906 & 11.651 & 10.837 &  0.203 &  0.161 &  0.129 &  13 &  13 &  13 & $  4.53$\\
  1898 & T 14                 &  166.037953 & $ -76.455369$ &  9.668 &  8.935 &  8.563 &  0.056 &  0.063 &  0.067 &  13 &  13 &  13 & $  1.47$\\
  2168 & \multicolumn{1}{c}{...} &  166.062871 & $ -76.110506$ & 14.640 & 14.091 & 13.932 &  0.117 &  0.107 &  0.096 &  12 &  12 &  12 & $  1.20$\\
\enddata
\end{deluxetable}
\clearpage

%\end{document}

%\documentclass[preprint]{aastex}
%\begin{document}

%\tablenum{4}
%\pagestyle{empty}
%\rotate

\begin{deluxetable}{rcc@{\extracolsep{20pt}}
      r@{\extracolsep{10pt}}r@{\extracolsep{10pt}}r@{\extracolsep{20pt}}
      c@{\extracolsep{10pt}}c@{\extracolsep{10pt}}c@{\extracolsep{10pt}} 
      c@{\extracolsep{0pt}}c@{\extracolsep{0pt}}c}
\tabletypesize{\scriptsize}
\tablewidth{490pt}
\tablecaption{New Candidate Members of Chamaeleon~I\label{tbl:candidates}}
\tablehead{
  \colhead{ID}        & 
  \multicolumn{2}{c}{Coordinates (J2000)} &
  \multicolumn{3}{c}{Mean Magnitudes}     &
  \multicolumn{3}{c}{Observed RMS}        &
  \colhead{Variable?} &
  \colhead{$JHK_s$}   &
  \colhead{Name}\\
  \cline{2-3}
  \cline{4-6}
  \cline{7-9}
  \colhead{}          &
  \colhead{$\alpha$}  &
  \colhead{$\delta$}  &
  \colhead{$J$}       &
  \colhead{$H$}       &
  \colhead{$K_s$}     &
  \colhead{$J$}       &
  \colhead{$H$}       &
  \colhead{$K_s$}     &
  \colhead{}          &
  \colhead{excess?}
}
\startdata
% 329 & 165.880201 & -77.817027 & 16.780 & 14.350 & 12.854 & 0.327 & 0.068 & 0.056 &   & Y & Cam2 10\\
% 926 & 165.939710 & -77.784451 & 16.882 & 14.390 & 12.832 & 0.183 & 0.035 & 0.027 &   & Y & Cam2 11\\
 1715 & 166.017772 & -76.659153 & 12.906 & 11.651 & 10.837 & 0.203 & 0.161 & 0.129 & Y & Y\\
%1959 & 166.042644 & -76.371713 &  9.952 &  9.848 &  9.728 & 0.018 & 0.024 & 0.021 &   & Y\\
 1982 & 166.044294 & -76.213627 & 13.141 & 12.522 & 12.125 & 0.016 & 0.044 & 0.015 &   & Y\\
 7869 & 166.636396 & -76.422525 & 14.270 & 13.597 & 13.153 & 0.049 & 0.065 & 0.043 &   & Y\\
 9484 & 166.799142 & -76.430582 & 13.667 & 12.904 & 12.466 & 0.066 & 0.054 & 0.041 & Y & \\
10193 & 166.874272 & -77.417151 & 12.163 & 11.353 & 10.958 & 0.121 & 0.112 & 0.111 & Y & \\
10862 & 166.944190 & -76.254861 & 13.972 & 13.031 & 12.374 & 0.045 & 0.047 & 0.040 &   & Y & GK 18\\
11564 & 167.014837 & -79.376292 & 14.459 & 13.483 & 12.265 & 0.052 & 0.043 & 0.041 &   & Y\\
13788 & 167.236477 & -77.724595 & 15.256 & 13.513 & 12.449 & 0.214 & 0.116 & 0.052 &   & Y & Cam2 35\\
\enddata
\end{deluxetable}
\clearpage

%\end{document}

%\documentclass[preprint]{aastex}
%\begin{document}

%\tablenum{5}

%\rotate

\begin{deluxetable}{lll@{\extracolsep{10pt}}
      r@{\extracolsep{8pt}}r@{\extracolsep{8pt}}r@{\extracolsep{10pt}}
      c@{\extracolsep{8pt}}c@{\extracolsep{8pt}}c@{\extracolsep{10pt}}
      c@{\extracolsep{-4pt}}c@{\extracolsep{-4pt}}c@{\extracolsep{0pt}}c
      @{\extracolsep{0pt}}c}
\tabletypesize{\scriptsize}
\tablewidth{540pt}
\tablecaption{Near-infrared Photometry for Previously Identified Chamaeleon~I Members\label{tbl:members}}
\tablehead{
  \colhead{Name}                          &
  \multicolumn{2}{c}{Coordinates (J2000)} &
  \multicolumn{3}{c}{Mean Magnitudes\tablenotemark{a}} &
  \multicolumn{3}{c}{RMS\tablenotemark{a}} &
  \multicolumn{3}{c}{N}                   &
  \colhead{Variability}                   &
  \colhead{YSO\tablenotemark{b}}\\
  \cline{2-3}
  \cline{4-6}
  \cline{7-9}
  \cline{10-12}
  \colhead{}          &
  \colhead{$\alpha$}  &
  \colhead{$\delta$}  &
  \colhead{$J$}       &
  \colhead{$H$}       &
  \colhead{$K_s$}     &
  \colhead{$J$}       &
  \colhead{$H$}       &
  \colhead{$K_s$}     &
  \colhead{$J$}       &
  \colhead{$H$}       &
  \colhead{$K_s$}     &
  \colhead{Index}     &
  \colhead{Properties}
}
\startdata
T 1             &     10:52:01.30 &     -77:09:50.1 & 13.942 & 13.333 & 13.112 & 0.027 & 0.037 & 0.037 &     1 &     1 &     1 & \multicolumn{1}{c}{...} & 0Y0000\\
IRAS 10529-7638 &      10:54:11.6 &       -76:54:33 & \multicolumn{1}{c}{...} & \multicolumn{1}{c}{...} & \multicolumn{1}{c}{...} & \multicolumn{1}{c}{...} & \multicolumn{1}{c}{...} & \multicolumn{1}{c}{...} & \multicolumn{1}{c}{...} & \multicolumn{1}{c}{...} & \multicolumn{1}{c}{...} & \multicolumn{1}{c}{...} & 00000Y\\
T 2             &     10:54:13.32 &     -77:55:09.2 & 12.251 & 11.699 & 11.477 & 0.023 & 0.026 & 0.029 &     1 &     1 &     1 & \multicolumn{1}{c}{...} & Y00000\\
T 3             &      10:55:59.9 &       -77:24:38 & \multicolumn{1}{c}{...} & \multicolumn{1}{c}{...} & \multicolumn{1}{c}{...} & \multicolumn{1}{c}{...} & \multicolumn{1}{c}{...} & \multicolumn{1}{c}{...} & \multicolumn{1}{c}{...} & \multicolumn{1}{c}{...} & \multicolumn{1}{c}{...} & \multicolumn{1}{c}{...} & YY000Y\\
T 4             &     10:56:30.43 &     -77:11:39.4 &  9.966 &  9.078 &  8.629 & 0.021 & 0.026 & 0.027 &     1 &     1 &     1 & \multicolumn{1}{c}{...} & YY0Y0Y\\
T 5             &     10:57:42.28 &     -76:59:35.7 & 10.431 &  9.496 &  9.196 & 0.022 & 0.026 & 0.025 &     1 &     1 &     1 & \multicolumn{1}{c}{...} & 0Y0000\\
CHXR 3          &     10:58:05.61 &     -77:28:24.0 &  8.507 &  7.675 &  7.392 & 0.025 & 0.034 & 0.023 &     1 &     1 &     1 & \multicolumn{1}{c}{...} & 0Y0Y0Y\\
T 6             &     10:58:16.84 &     -77:17:17.1 &  9.270 &  8.414 &  7.783 & 0.025 & 0.037 & 0.026 &     1 &     1 &     1 & \multicolumn{1}{c}{...} & YY0Y0Y\\
\enddata
\tablenotetext{a}{For sources with $J$, $H$, and $K_s$ photometry, N=1 in all three
                  bands, but otherwise no tabulated Stetson index, the photometry is
                  from the Version 2 processing of the 2MASS Working Database and
                  the RMS represents the photometric uncertainty from the 2MASS
                  processing pipeline.}
\tablenotetext{b}{Six character flag indicating the properties of that star. References provided in the text.}
\tablenotetext{\ }{First character : Variable identified from this study or General Catalog of Variable Stars}
\tablenotetext{\ }{Second character: Contains H$\alpha$ in emission}
\tablenotetext{\ }{Third character : Contains lithium in absorption}
\tablenotetext{\ }{Fourth character: X-ray source}
\tablenotetext{\ }{Fifth character : Contains an infrared excess}
\tablenotetext{\ }{Sixth character : Far-infrared source}
\end{deluxetable}
\clearpage

%\end{document}

\begin{table}
%\dummytable\label{tbl:coords}
%\dummytable\label{tbl:obslog}
%\dummytable\label{tbl:variables}
%\dummytable\label{tbl:candidates}
%\dummytable\label{tbl:members}
\dummytable\label{tbl:crossids}
\end{table}

\clearpage

\clearpage

%\clearpage

\begin{figure}
\insertplot{figure3.ps}{7.8}{8.4}{0.0}{0.8}{0.75}{1}
\caption{
  Photometric data for star 11344 (also known as T~29 and Sz~22) that 
  illustrates the data obtained for this study. The left and middle panels 
  show the $J$, $H$, \KB, $J-H$ and \HK\ light curves. The vertical bars 
  through the data points represent the $\pm 1\sigma$ photometric 
  uncertainties. The right panels show the \KB\ vs. \HK\ color-magnitude 
  diagram and the $J-H$ vs. \HK\ color-color diagrams for each data point in 
  the time series, where the dotted line represents the interstellar reddening 
  vector from \citet{Cohen81} transformed into the 2MASS photometric system 
  \citep{Carp01}. The uncertainties in the stellar colors have been omitted
  for clarity. The solid line in the color-magnitude diagram is the 
  2~Myr pre-main-sequence isochrone from \citet{DM97} for stellar masses 
  between 0.08\msun\ and 3\msun. The solid curves in the color-color diagram 
  are the loci of red giant and main-sequence stars from \citet{BB88} in the 
  2MASS color system.
  \label{fig:variable}
}
\end{figure}
\clearpage

\clearpage

\clearpage

\clearpage

\clearpage

\clearpage

\clearpage

\end{document}